\documentclass[fleqn,usenatbib]{mnras}

\usepackage{newtxtext,newtxmath}

\usepackage[T1]{fontenc}

\DeclareRobustCommand{\VAN}[3]{#2}
\let\VANthebibliography\thebibliography
\def\thebibliography{\DeclareRobustCommand{\VAN}[3]{##3}\VANthebibliography}

\usepackage{graphicx}	
\usepackage{amsmath}
\usepackage{xcolor}
\usepackage{tabularx}
\usepackage{footnote}
\usepackage{pdflscape}
\usepackage{url}
\usepackage{booktabs,caption}
\usepackage[flushleft]{threeparttable}
\usepackage{subfig}

\title[The Volumetric Extended-Schmidt Law]{The Volumetric Extended-Schmidt Law: A Unity Slope}

\author[K. Du et al.]{
Kaiyi Du,$^{1,2}$
Yong Shi,$^{1,2}$\thanks{Email: yong@nju.edu.cn}
Zhi-Yu Zhang,$^{1,2}$
Qiusheng Gu,$^{1,2}$
Tao Wang,$^{1,2}$
Junzhi Wang,$^{3}$
Xin Li$^{1,2}$
\newauthor
and Sai Zhai$^{1,2}$
\\
$^{1}$School of Astronomy and Space Science, Nanjing University, Nanjing 210093, People’s Republic of China \\
$^{2}$Key Laboratory of Modern Astronomy and Astrophysics (Nanjing University), Ministry of Education, Nanjing 210093, People’s Republic of China \\
$^{3}$School of Physical Science and Technology, Guangxi University, Nanning 530004, People’s Republic of China \\
}

\date{Accepted XXX. Received YYY; in original form ZZZ}

\pubyear{2022}

\begin{document}
\label{firstpage}
\pagerange{\pageref{firstpage}--\pageref{lastpage}}
\maketitle

\begin{abstract}

We investigate the extended-Schmidt (ES) law in volume densities ($\rho_{\rm SFR}$ $\propto$ $(\rho_{\rm gas}\rho_{\rm star}^{0.5})^{\alpha^{\rm VES}}$) for spatially-resolved regions in spiral, dwarf, and ultra-diffuse galaxies (UDGs), and compare to the volumetric Kennicutt-Schmidt (KS) law ($\rho_{\rm SFR}$ $\propto$ $\rho_{\rm gas}^{\alpha^{\rm VKS}}$). 
We first characterize these star formation laws in individual galaxies using a sample of 11 spirals, finding median slopes $\alpha^{\rm VES}$=0.98 and $\alpha^{\rm VKS}$=1.42, with a galaxy-to-galaxy rms fluctuation that is substantially smaller for the volumetric ES law (0.18 vs 0.41). By combining all regions in spirals with those in additional 13 dwarfs and one UDG into one single dataset, it is found that the rms scatter of the volumetric ES law at given x-axis is 0.25 dex, also smaller than that of the volumetric KS law (0.34 dex). At the extremely low gas density regime as offered by the UDG, the volumetric KS law breaks down but the volumetric ES law still holds. On the other hand, as compared to the surface density ES law, the volumetric ES law instead has a slightly larger rms scatter, consistent with the scenario that the ES law has an intrinsic slope of $\alpha^{\rm VES} \equiv$1 but the additional observational error of the scale height increases the uncertainty of the volume density. The unity slope of the ES law implies that the star formation efficiency (=$\rho_{\rm SFR}$/$\rho_{\rm gas}$) is regulated by the quantity that is related to the $\rho_{\rm star}^{0.5}$.

\end{abstract}




\section{Introduction}\label{sec.intro}

Stars are building blocks of galaxies. The star formation process is complex while crucial to galaxy formation and evolution. Stars form in clouds of cold gas \citep{Bergin2007}. Supernova explosion and stellar wind from massive stars result in chemical enrichment and energy feedback to the surrounding interstellar medium (ISM) \citep{Fierlinger2016}, regulating the subsequent formation of stars. Understanding the gas-star formation cycle is crucial for understanding the life cycle of galaxies and the whole cosmic ecosystem. Despite lacking a full understanding of the star formation process, empirical scaling relations between star formation rate (SFR) and cold gas characterize the universal trends from gas to new stars across cosmic time and provide useful recipes for theoretical models and simulations \citep{Semenov2018}.

The star formation relation was first proposed by \cite{Schmidt1959} in volume densities:
\begin{equation}
    \rho_{\rm SFR} \propto \rho_{\rm gas}^{\alpha},
    \label{eq.1}
\end{equation}
where the SFR is presumed to depend only on the total gas density, and $\alpha = 1-2$. The Schmidt law has been used to describe star formation on large-scale in galaxies and used as a recipe for star formation in cosmological simulations. Other early studies (e.g \citealt{Berkhuijsen1977,Freedman1984}) that measure the form of the Schmidt law show a large dispersion in slope with $0<\alpha<4$. 

\cite{Kennicutt1989,Kennicutt1998} have derived this relationship in terms of the surface density between the SFR and total cold gas (HI+H$_{2}$) for 61 nearby spiral galaxies and 36 nuclear starburst regions that span a dynamic range of seven orders of magnitude: 
\begin{equation}
    \Sigma_{\rm SFR}\propto\Sigma_{\rm gas}^{N},
    \label{eq.2}
\end{equation}
which is the so-called Kennicutt-Schmidt (KS) law with $N=1.4 \pm 0.15$. However, this relationship breaks down in low surface density environments, such as low-surface-brightness (LSB) galaxies and outer disks of galaxies, whose $\Sigma_{\rm SFR}$ is below about 10$^{-3}$ M$_{\odot}$/yr/kpc$^{2}$ \citep{Kennicutt1998, Wyder2009, Shi2018}. It implies that there might exist additional parameters besides the total gas mass that may regulate ongoing star formation. One modified version of the KS law is the Silk-Elmegreen relationship \citep{Elmegreen1997,Silk1997} involving the dynamical orbital timescale ($\Sigma_{\rm SFR}\propto\Sigma_{\rm gas}/t_{\rm dyn}$), which implies that global dynamical features like spiral arms might convert a constant fraction of gas into stars, but this relationship fails for outer disks of dwarfs \citep{Shi2018}. Another one is the extended-Schmidt (ES) law that includes the stellar mass surface density in the correlation ($\Sigma_{\rm SFR}\propto\Sigma_{\rm gas}\Sigma_{\rm star}^{0.5}$) \citep{Shi2011,Shi2018}, which reveals an important role for existing stars in regulating star formation through their gravitational effect on the mid-plane pressure. The ES law holds down to very low SFR surface densities including outer disks of dwarfs as well as ultra diffuse galaxies (\citealt{Shi2021}; Zhai et al. 2022, submitted). Also, from the KS law to the ES law, the offset of extremely metal-poor star-forming galaxies/regions is significantly reduced \citep{Roychowdhury2017,Shi2018}. 

While the surface densities are easy to measure, the volume densities are more physical quantities because the gas disc is flaring and will result in non-negligible projection effects. Besides, the star formation law applied to simulations is generally based on volume densities \citep{Schaye2008,Semenov2018}. \citet{Krumholz2012} also emphasizes the importance of volumetric star formation law by comparing the observations and theoretical models. \cite{Yim2011,Yim2014,Yim2020} have measured the gas disk thickness and radial profile of mid-plane volume densities for five edge-on galaxies and investigated the volumetric KS law. Using the Poisson equation under hydrostatic equilibrium, \citet{Bacchini2019,Bacchini2019a,Bacchini2020} confirmed that the volumetric KS law shows smaller scatters than the surface-density one and less deviations toward the low density end.

Given the good performance of the ES law in surface densities, in this paper we study the ES law in volume densities for 11 nearby spiral galaxies, 13 nearby dwarf galaxies and one ultra-diffuse galaxy (UDG) AGC~242019 in order to better understand the role of stellar mass in regulating star formation in a 3-D galaxy. We present the data and the measurements of physical quantities in Section~\ref{sec.sample}. In Section~\ref{sec.model}, we present the general model and the method to calculate the scale heights of each baryonic component. Section~\ref{sec.result} shows the main results of this work and compare the volumetric ES law to the volumetric KS law. In Section~\ref{sec.discusssion}, we discuss the factors that may influence the calculation of volume densities and the physical implication of the volumetric ES law. Section~\ref{sec.conclusion} gives a conclusion for our work.

\section{Data and Measurements}\label{sec.sample}

\begin{table*}
  \begin{threeparttable}
    \caption{Basic properties of the sample galaxies.}
    \label{tab.1}
     \begin{tabular}{cccccccc}
        \toprule
        \begin{tabular}[c]{@{}c@{}}Galaxy\\ \\(1)\end{tabular}& \begin{tabular}[c]{@{}c@{}}RA (J2000)\\(h m s)\\(2)\end{tabular} & \begin{tabular}[c]{@{}c@{}}DEC (J2000)\\($^\circ$ $'$ $''$ )\\(3)\end{tabular} & \begin{tabular}[c]{@{}c@{}}Morphology\\ \\ (4)\end{tabular}& \begin{tabular}[c]{@{}c@{}}Distance\\(Mpc)\\(5)\end{tabular} & \begin{tabular}[c]{@{}c@{}}i\\(deg)\\(6)\end{tabular} & \begin{tabular}[c]{@{}c@{}}PA\\(deg)\\(7)\end{tabular} & \begin{tabular}[c]{@{}c@{}}E(B-V)\\ \\(8)\end{tabular} \\
        \midrule
        NGC~628 & 01 36 41.8 & +15 47 00 & SAc   & 7.30  & 7.00  & 20.00 & 0.068 \\
        NGC~925 & 02 27 16.9 & +33 34 45 & SABd  & 9.20  & 58.00 & 287.00 & 0.073 \\
        NGC~2403 & 07 36 51.4 & +65 36 09 & Scd   & 3.16  & 61.00 & 124.50 & 0.038 \\
        NGC~2841 & 09 22 02.6 & +50 58 35 & SAb   & 14.10 & 74.00 & 153.00 & 0.015 \\
        NGC~2976 & 09 47 15.5 & +67 54 59 & SAc   & 3.58  & 61.00 & 334.50 & 0.068 \\
        NGC~3198 & 10 19 55.0 & +45 32 59 & SBc   & 13.80 & 71.50 & 216.00 & 0.012 \\
        NGC~3521 & 11 05 48.6 & -00 02 09 & SABbc & 10.70 & 73.00 & 340.00 & 0.055 \\
        NGC~4736 & 12 50 53.0 & +41 07 13 & SAab  & 4.70  & 41.00 & 296.00 & 0.017 \\
        NGC~5055 & 13 15 49.2 & +42 01 45 & SAbc  & 10.10 & 59.00 & 102.00 & 0.017 \\
        NGC~6946 & 20 34 52.2 & +60 09 14 & SABcd & 5.90  & 33.00 & 243.00 & 0.328 \\
        NGC~7331 & 22 37 04.1 & +34 24 57 & SAb   & 14.70 & 76.00 & 168.00 & 0.087 \\
        \midrule
        CVnIdwA & 12 38 39.2 & +32 45 41.0 & dIrr & 3.60 & 66.50 & 48.40 & 0.015 \\
        DDO~50 & 08 19 03.7 & +70 43 24.6 & dIrr & 3.40 & 49.70 & 175.70 & 0.031 \\
        DDO~52 & 08 28 28.4 & +41 51 26.5 & dIrr & 10.30 & 43.00 & 8.20 & 0.034 \\
        DDO~53 & 08 34 06.4 & +66 10 47.9 & dIrr & 2.60 & 27.00 & 131.60 & 0.036 \\
        DDO~87 & 10 49 34.9 & +65 31 47.9 & dIrr & 7.70 & 55.50 & 235.10 & 0.010 \\
        DDO~101 & 11 55 39.1 & +31 31 9.9 & dIrr & 6.40 & 51.00 & 287.40 & 0.021 \\
        DDO~126 & 12 27 06.6 & +37 08 15.9 & dIrr & 4.90 & 65.00 & 138.00 & 0.014 \\
        DDO~133 & 12 32 55.2 & +31 32 19.1 & dIrr & 3.50 & 43.40 & 359.60 & 0.015 \\
        DDO~154 & 12 54 05.7 & +27 09 09.9 & dIrr & 3.70 & 68.20 & 226.30 & 0.009\\
        DDO~168 & 13 14 27.3 & +45 55 37.3 & dIrr & 4.30 & 46.50 & 275.50 & 0.014\\
        DDO~216 & 23 28 34.7 & +14 44 56.2 & dIrr & 1.1 & 63.70 & 133.60 & 0.065\\
        NGC~2366 & 07 28 53.4 & +69 12 49.6 & dIrr & 3.40 & 63.00 & 38.70 & 0.035 \\
        WLM & 00 01 59.9 & -15 27 57.2 & dIrr & 1.00 & 74.00 & 174.50 & 0.036 \\
        \midrule
        AGC~242019 & 14 33 53.4 & +01 29 12.5 & dIrr & 30.80 & 73.00 & 2.00 & 0.045 \\
        \bottomrule
     \end{tabular}
    \begin{tablenotes}
      \small
      \item Note. The basic properties of our sample are taken from \cite{Shi2011}, \cite{Oh2015}, \cite{Bacchini2019} and \cite{Shi2021}. Col.(1): the name of the galaxies; Col.(2)-(3): coordinates of the galaxies; Col.(4): the morphologies of galaxies taken from the NASA/IPAC Extragalactic Database (NED); Col.(5)-(7): the distances, inclinations and position angle of the galaxies; Col.(8): the foreground reddening taken from NED \citep{Schlafly2011}. 
    \end{tablenotes}
  \end{threeparttable}
\end{table*}

To study the ES law in volume densities, we select a sample of 11 spiral galaxies, 13 dwarf galaxies and one HI-rich UDG AGC~242019, all of which have available data of atomic gas, molecular gas, stellar mass, and SFR.  The rotation curves, dark matter halo profiles and vertical velocity dispersion of gas components are available in the literature or estimated by 3D-Barolo fitting \citep{DiTeodoro2015} in this work. 

Spiral galaxies were selected by cross-matching the HI Nearby Galaxy Survey (THINGS, \citealt{Walter2008}), and the HERA CO line extragalactic survey (HERACLES, \citealt{Leroy2005}) because they have sufficient spatial resolution. Our dwarf galaxies were from the Local Irregulars That Trace Luminosity Extremes The HI Nearby Galaxy Survey (LITTLE THINGS, \citealt{Hunter2012}) with rotation curves extracted by \cite{Oh2015}. We note that only four dwarfs in LITTLE THINGS have CO observations, and even these objects have low molecular fraction with $f_{\rm H_2} = \frac{M_{\rm H_2}}{M_{\rm H_2}+M_{\rm HI}} \sim 0.23$ \citep{Hunter2019}. Hence, we ignore the contribution of molecular gas in dwarf galaxies and take the atomic gas masses as the total gas masses. UDGs are as faint as dwarf galaxies but their size are comparable to spiral galaxies. As AGC~242019 has high S/N HI 21 cm observations from the Karl G. Jansky Very Large Array (VLA) with rotation curve and dark matter halo profile well measured by \cite{Shi2021}, we include it in our sample too. We also ignore the molecular gas in AGC~242019 because UDGs have little molecular gas \citep{Wang2020}. The spatially-resolved measurements of AGC~242019 are detailed in Zhai et al. (2022, submitted).

For spiral galaxies, the total gas masses are the sum of atomic gas and molecular gas ($\Sigma_{\rm gas}=\Sigma_{\rm HI}+\Sigma_{\rm H_2}$), which are derived from the THINGS 21 cm emission maps and HERACELS CO ($J=2-1$) emission maps, following the equations in \citet{Walter2008} and \citet{Leroy2013}, respectively:
\begin{equation}
    \Sigma_{\rm HI}=1.20\times10^{4}{\rm cos}(i)(1+z)^{3}\frac{{\rm arcsec}^{2}}{b_{\rm maj}\times b_{\rm min}}S_{\rm HI}\Delta v,
    \label{eq.3}
\end{equation}
where $\Sigma_{\rm HI}$ is in M$_{\odot}$/pc$^{2}$ with the lower-limit of surface density $\sim$ 1.5\ M$_{\odot}$/pc$^{2}$. $i$ is the inclination angle of the galaxy, $z$ is the redshift, $b_{\rm maj}$ and $b_{\rm min}$ are the major and minor beam sizes in arcsec of the systematic beam of HI map, $S_{\rm HI}\Delta v$ is in ${\rm Jy\ km\ s}^{-1}$beam$^{-1}$.
\begin{equation}
    \Sigma_{\rm H_2}=6.3\ {\rm cos}(i)\ (\frac{0.7}{R_{21}})\ (\frac{\alpha_{\rm CO}^{1-0}}{4.35})\ I_{\rm CO},
    \label{eq.4}
\end{equation}
where the $\Sigma_{\rm H_2}$ is in M$_{\odot}$/pc$^{2}$, $R_{21}$ is the line ratio of CO ($J=2-1$) to CO ($J=1-0$), $\alpha_{\rm CO}^{1-0}$ is the conversion factor of CO ($J=1-0$)-to-H$_{2}$, and $I_{\rm CO}$ is in ${\rm K\ km\ s}^{-1}$. Here, we adopt the $R_{21} = 0.7$ \citep{Dame2001,Leroy2013} and a Galactic conversion factor $\alpha_{\rm CO}^{1-0} = 4.35\ {\rm M}_{\odot}\ {\rm pc}^{-2}({\rm K\ km\ s}^{-1})^{-1}$ \citep{Bolatto2013}. Eq.~\ref{eq.1} and Eq.~\ref{eq.2} have been multiplied by a factor of 1.36 to account for the helium. For the measurement of molecular gas, we give a lower-limit of surface density $\sim 3$ M$_{\odot}\ /{\rm pc}^{2}$ due to the sensitivity of HERACLES observations \citep{Bigiel2008}. For regions with $\Sigma_{\rm H_2}$ below this limit, we ignore their molecular gas masses. The SFRs are obtained by combining the infrared data from the SIRTF Nearby Galaxies Survey (SINGS, \cite{Kennicutt2003}) and the far-UV data from GALEX archive \citep{GildePaz2007}, using the following formula \citep{Leroy2008}:
\begin{equation}
    \Sigma_{\rm SFR}=8.1\times 10^{-2}\ {\rm cos}(i)\ I_{\rm FUV}+3.2\times10^{-3}\ {\rm cos}(i)\ I_{\rm 24\mu m},
    \label{eq.5}
\end{equation}
where the $\Sigma_{\rm SFR}$ is in ${\rm M}_{\odot}/{\rm yr}/{\rm kpc}^{2}$ with the 3-$\sigma$ lower-limit of $10^{-4}\ {\rm M}_{\odot}/{\rm yr}/{\rm kpc}^{2}$ \citep{Shi2018}, $I_{\rm FUV}$ and $I_{\rm 24\mu m}$ are in $\rm MJy/sr$. The FUV flux densities are corrected for Galactic extinction using the Galactic extinction curve derived by \cite{Cardelli1989} ($R_{V}=A_{V}/E(B-V)=3.1)$ that gives $A_{\rm FUV}=7.9E(B-V)$ \citep{Hao2011}).
The stellar masses are estimated based on SINGS $3.6\mu {\rm m}$ data using the following equation \citep{Leroy2008}:
\begin{equation}
    \Sigma_{\rm star}=140\ {\rm cos}(i)\ I_{\rm 2mass-K}=280\ {\rm cos}(i)\ I_{3.6\rm \mu m},
    \label{eq.6}
\end{equation}
where the $I_{3.6\rm \mu m}$ is in $\rm MJy/sr$. 

In this work, we also need the profile of dark matter halo in order to calculate the scale heights of gaseous disks. We cross-match all the surveys mentioned above with the sample in \citet{deBlok2008} and \citet{Oh2015} that obtained dark matter halo profiles for a sample of 10 spiral galaxies and 13 dwarf galaxies. There is a face-on galaxy NGC~628 whose profiles of its dark matter halo and rotation curve were retrieved from \citet{Aniyan2018}. Finally, we have a final sample of 25 galaxies, and the basic information and the dark matter profile parameters are listed in Table~\ref{tab.1} and Table~\ref{tab.2}, respectively.

To have a uniform spatial resolution across the wavelength, we convolve all images with kernels from \citet{Aniano2011} to have resolution beams of $14''\times14''$. We then carry out spatially-resolved measurements of gas masses, stellar masses, and SFRs for our sample following \citet{Shi2011}, with 1000 pc $\times$ 1000 pc apertures across disks of galaxies out to regions with $\Sigma_{\rm SFR} > 10^{-4}\ {\rm M}_{\odot}/{\rm yr}/{\rm kpc}^{2}$ and $\Sigma_{\rm HI} > 1.5\ {\rm M}_{\odot}/{\rm pc}^{2}$.


\section{Method}\label{sec.model}

\subsection{Model}
We assume that each galaxy is an axisymmetric rotating disk in hydrostatic equilibrium under the gravitational potential. We use a three-component galactic disk model of gravitationally coupled atomic gas, molecular gas, and stars, subjected to the external gravitational field of dark matter halo.

The Poisson equation in cylindrical coordinates for an axisymmetric system is \citep{Banerjee2011}: 
\begin{equation}
     \frac{1}{R} \frac{\partial}{\partial R}(R \frac{\partial \Phi_{\rm total}}{\partial R})+\frac{\partial^{2}\Phi_{\rm total}}{\partial^{2}z} = 4 \pi G (\sum^{3}_{i=1}\rho_{i}+\rho_{\rm h}),
     \label{eq.7}
\end{equation}
where the $\Phi_{\rm total}$ is the gravitational potential of the galactic disk and the dark matter halo, $\rho_{i}$ with $i =$ 1 to 3 represents the volume density for each baryonic component (atomic gas, molecular gas and stars) and $\rho_{\rm h}$ represents the volume density of dark matter halo.

Along z-direction, each component follows the equation of hydrostatic equilibrium \citep{Rohlfs1977}:
\begin{equation}
\frac{\partial}{\partial z}(\rho_{i}\langle \sigma^{2}_{z}\rangle _{i}) +      \rho_{i} \frac{\partial \Phi_{\rm total}}{\partial z} = 0,
     \label{eq.8}
\end{equation}
where the $\langle \sigma^{2}_{z}\rangle _{i}$ is the mean square velocity of $i^{th}$ component.

Using Eq.~\ref{eq.8} and assuming vertically isothermal conditions for each disk component, we can rewrite Eq.~\ref{eq.7} as:
\begin{equation}
\begin{split}
\langle \sigma^{2}_{z} \rangle _{i} \frac{\partial}{\partial z}(\frac{1}{\rho_{i}})=&-4\pi G(\sum^{3}_{i=1}\rho_{i}+\rho_{\rm h}) +\frac{1}{R} \frac{\partial}{\partial R} (R\frac{\partial \Phi_{\rm total}}{\partial R}).
\label{eq.9}
\end{split}
\end{equation}

The radial derivative term of the potential on the right-hand side can be estimated from the rotation curve of the galaxy:
\begin{equation}
    (R\frac{\partial \Phi_{\rm total}}{\partial R})_{R,z}=(v^2_{\rm rot})_{R,z},
    \label{eq.10}
\end{equation}
where the $(v_{\rm rot})_{R,z}$ is the rotational velocity at any $R$ and $z$. Since the observed rotation curve only gives the intensity-weighted average of $(v_{\rm rot})_{R,z}$ along z-direction and we cannot accurately measure the $v_{\rm rot}$ at each $z$ on a given $R$, we simply assume that the rotational velocity does not change along the z-direction and is approximately equal to the observed rotational velocity. Then Eq.~\ref{eq.9} can be written as:
\begin{equation}
\begin{split}
\langle \sigma^{2}_{z} \rangle _{i} \frac{\partial}{\partial z}(\frac{1}{\rho_{i}}\frac{\partial\rho_i}{\partial z})=&-4\pi G(\sum^{3}_{i=1}\rho_{i}+\rho_{\rm h}) +\frac{1}{R} \frac{\partial}{\partial R} (v^2_{\rm rot}),
\label{eq.11}
\end{split}
\end{equation}
which represents three coupled second-order ordinary differential equations.

\begin{figure*}
  \centering
  \includegraphics[scale=0.5]{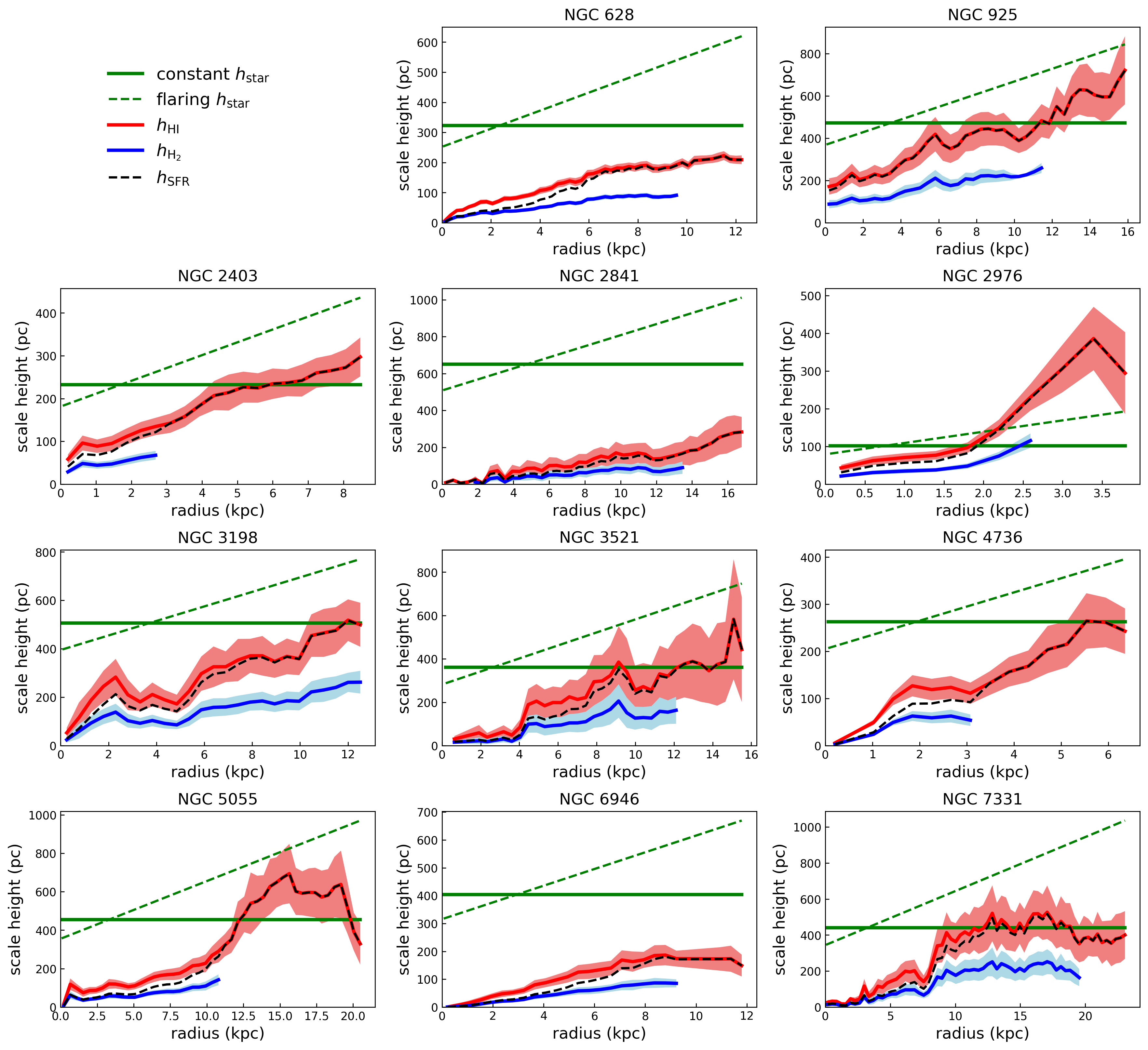}
    \caption{The radial profiles of scale heights of atomic gas, molecular gas and stellar disks are plotted for each galaxy. The red lines represent the scale height of HI, the blue lines represent the scale height of H$_2$, while the black dashed lines are the scale height of SFR estimated using Eq.~\ref{eq.11}. The green solid and dashed lines represent the constant (listed in Table~\ref{tab.2}) and flaring ($h(R)=\frac{30\ {\rm pc}}{\rm kpc}\times (R-l_{\rm star})+l_{\rm star}/7.3$) scale height of stellar disks, respectively.}
    \label{fig.1}
\end{figure*}

\begin{figure*}
  \ContinuedFloat
  \centering
  \includegraphics[scale=0.5]{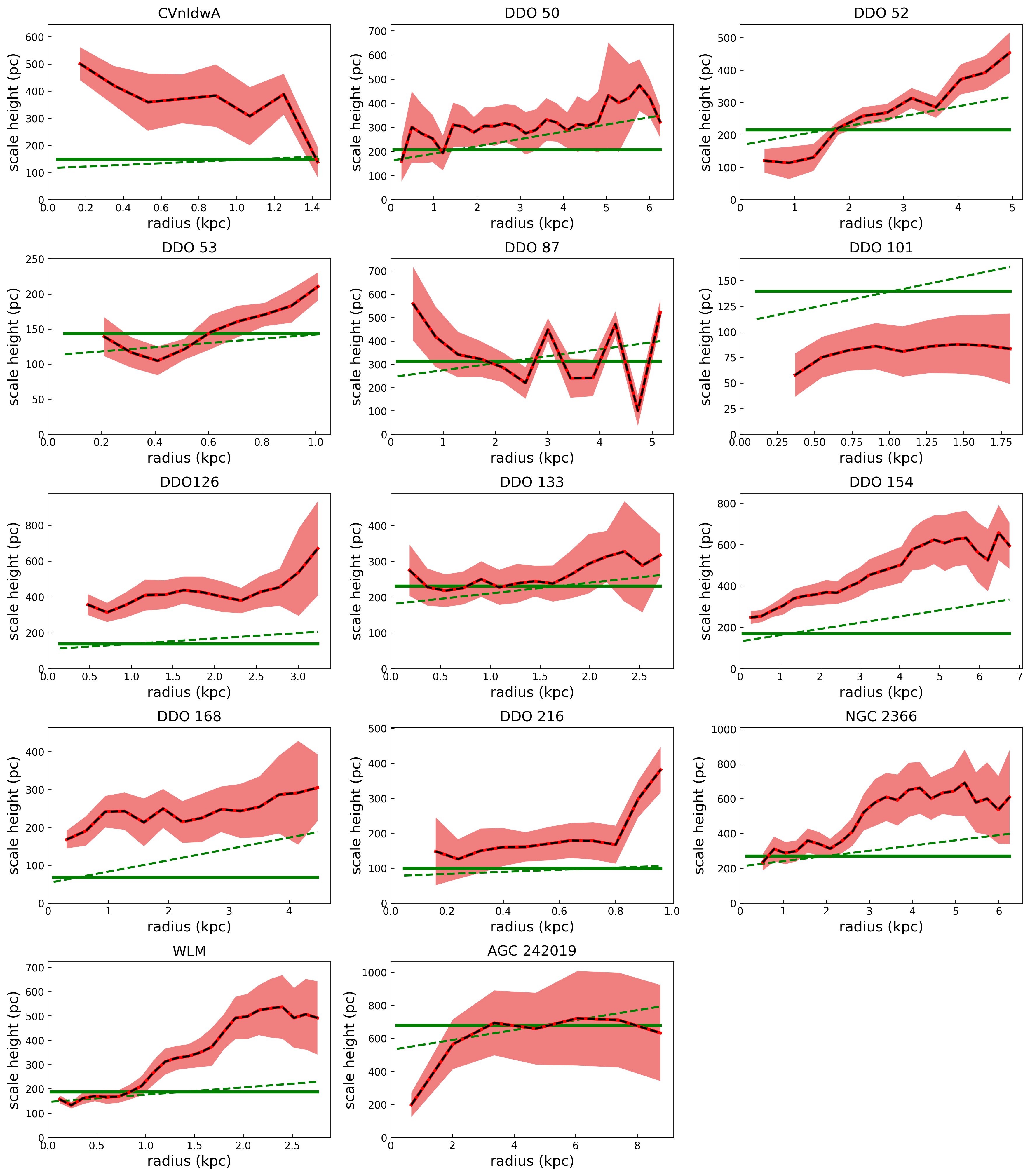}
    \caption{Continued.}
    \label{fig.1}
\end{figure*}

\subsection{The velocity dispersion of gas components}
To solve Eq.~\ref{eq.11}, we need measurements of the vertical velocity dispersion on the right-hand side of Eq.~\ref{eq.11} for each component. THINGS and LITTLE THINGS provides high-spatial-resolution and high-velocity-resolution HI observations for nearby spiral galaxies, from which $\sigma_z$ of HI cam be determined via tilted-ring modelling. In this study, we adopt the radial profiles of HI velocity dispersion determined by \cite{Bacchini2019}, \cite{Iorio2017} and \cite{Shi2021} and 3D-Barolo modelling \citep{DiTeodoro2015} (assuming velocity dispersion is isotropic). However, it is hard to measure $\sigma_z$ for stars and molecular gas. For the stellar component, we assume the vertical distribution and adopt the empirical value of scale height from observations of edge-on galaxies. By assuming a stellar vertical distribution, we can avoid measuring $\sigma_z$ of stars to solve Eq.~\ref{eq.11}. As for molecular gas, the observed $\sigma_z$ is also hard to get because there are few high-resolution observations available for our sample. Thus we follow the assumption in \cite{Bacchini2019} of $\sigma_{\rm HI}/\sigma_{\rm H_2} \approx 2$, which is based on studies of CO velocity dispersion radial profiles \citep{CalduPrimo2013,Mogotsi2016,Marasco2017}.

\subsection{The stellar component}

We fit the stellar surface brightness profile of each galaxy and obtain the disk scale length with S\`{e}rsic profiles. Since most of our spiral galaxies have bulges, we perform one-dimensional bulge-to-disk decomposition  by dividing the stellar component into two parts: bulge ($\rho_{\rm b}$) and disk ($\rho_{\rm d}$). The results are shown in Table~\ref{tab.2}.
Because we can not directly measure the vertical velocity dispersion of stellar components, we assume a sech$^{2}$ vertical mass-distribution for stellar disk \citep{vanderKruit1981a}:
\begin{equation}
    \rho_{\rm d}(z) = \rho_{\rm d,0}{\rm sech}^{2}(\frac{z}{h_{\rm star}}),
    \label{eq.12}
\end{equation}
where the $\rho_{\rm d,0}$ is the volume density of stellar disk on the mid-plane and $h_{\rm star}$ is the stellar scale height at given radius. We adopt $h_{\rm star} = l_{\rm star} / (7.3\pm 2.2)$, where $l_{\rm star}$ is the stellar scale length \citep{Kregel2002,Banerjee2011}. With this assumption of stellar component, Eq.~\ref{eq.11} can be reduced to two ordinary differential equations of HI and H$_{2}$ components.

The mass distribution of bulges are modelled by using an ellipsoid with exponential profile \citep{deBlok2008}:
\begin{equation}
    \rho_{\rm b}(r) = \rho_{\rm b,0}\ {\rm exp}(-\frac{r}{r_{\rm b}}),
    \label{eq.13}
\end{equation}
where $\rho_{\rm b,0}$ is the central volume density, $r_{\rm b}$ is the scale-radius of bulge, and $r=\sqrt{R^2+(z/q_{\rm b})^2}$ at given $R$ and $z$ where $q_{\rm b}$ is the axis-ratio of the bulge. \cite{Noordermeer2007} derived an average intrinsic axis-ratio $q_{\rm b}=0.55\pm 0.12$ for 21 early-type disk galaxies. The median value of $q_{\rm b}$ in \cite{Mosenkov2010a} is $\sim 0.63$ for 2MASS edge-on spiral galaxies, and \cite{Pastrav2013} gave an average axis-ratio of 0.69 for CALIFA unbarred galaxies. Here, we thus adopt $q_{\rm b}=0.6$ for all bulges in our sample. The $\rho_{\rm b,0}$ and $r_{\rm b}$ can be obtained by fitting the exponential profile to the surface density of the bulge component.

\begin{table*}
  \begin{threeparttable}
    \caption{Parameters of surface brightness profiles and dark matter halo profiles for the sample.}
    \label{tab.2}
     \begin{tabular}{c|ccccc|ccc}
        \toprule
        \begin{tabular}[c]{@{}c@{}}Galaxy\\ \\(1)\end{tabular} & \begin{tabular}[c]{@{}c@{}}$\Sigma_{\rm star,0}$ \\M$_{\odot}$pc$^{-2}$\\(2)\end{tabular} & \begin{tabular}[c]{@{}c@{}}$l_{\rm star}$\\pc\\(3)\end{tabular} & \begin{tabular}[c]{@{}c@{}}$h_{\rm star}$\\pc\\(4)\end{tabular} & \begin{tabular}[c]{@{}c@{}}$\rho_{\rm b,0}$\\M$_{\odot}$pc$^{-3}$\\(5)\end{tabular} & \begin{tabular}[c]{@{}c@{}}$r_{\rm b}$\\pc\\(6)\end{tabular}& \begin{tabular}[c]{@{}c@{}}DM halo profile\\Type\\(7)\end{tabular} & \begin{tabular}[c]{@{}c@{}}c ; $\rho_0$\\ - ;  $10^{-3}$ M$_{\odot}$pc$^{-3}$\\(8)\end{tabular} & \begin{tabular}[c]{@{}c@{}}$V_{\rm 200}$ ; $r_{\rm c}$\\km/s ; kpc\\(9)\end{tabular}\\
        \midrule
        NGC~628 & 377.9 & 2357.0 & 322.9 & 2.49 & 254.0 & ISO & 642.90 & 0.69 \\
        NGC~925 & 119.8 & 3442.8 & 471.6 & - & - & ISO & 5.90 & 9.67 \\
        NGC~2403 & 293.0 & 1695.3 & 232.2 & - & - & NFW & 12.40 & 101.70 \\
        NGC~2841 & 454.1 & 4747.2 & 650.3 & 1.35 & 660.4 & ISO & 3215.30 & 0.62 \\
        NGC~2976 & 430.0 & 740.0 & 101.3 & - & - & ISO & 35.50 & 5.09 \\
        NGC~3198 & 179.5 & 3692.1 & 505.8 & 0.19 & 516.2 & ISO & 44.00 & 2.82\\
        NGC~3521 & 1192.2 & 2634.8 & 360.9 & 1.79 & 453.2 & ISO & 73.00 & 2.50 \\
        NGC~4736 & 493.2 & 1917.5 & 262.7 & 60.98 & 234.4 & NFW & 63.50 & 42.40 \\
        NGC~5055 & 876.0 & 3320.5 & 454.9 & 5.34 & 414.8 & ISO & 4.80 & 11.73 \\
        NGC~6946 & 668.5 & 2950.3 & 404.1 & 18.92 & 147.5 & ISO & 45.70 & 3.62 \\
        NGC~7331 & 1180.9 & 3216.4 & 440.6 & 1.14 & 681.4 & NFW & 4.90 & 200.00 \\
        \midrule
        CVnIdwA & 5.2 & 1084.3 & 148.5 & - & - & ISO & 8.19 & 2.01\\
        DDO~50 & 29.4 & 1507.0 & 206.4 & - & - & ISO & 379.64 & 0.15 \\
        DDO~52 & 11.4 & 1570.5 & 215.1 & - & - & ISO & 48.81 & 1.33 \\
        DDO~53 & 6.5 & 1046.2 & 143.3 & - & - & ISO & 25.10 & 2.22 \\
        DDO~87 & 7.5 & 2281.1 & 312.5 & - & - & ISO & 13.91 & 2.46 \\
        DDO~101 & 22.1 & 1018.7 & 139.5 & - & - & ISO & 849.14 & 0.32 \\
        DDO~126 & 10.7 & 1007.4 & 138.0 & - & - & ISO & 21.59 & 1.33 \\
        DDO~133 & 10.1 & 1683.2 & 230.6 & - & - & ISO & 73.69 & 0.83 \\
        DDO~154 & 4.7 & 1229.6 & 168.4 & - & - & ISO & 53.21 & 0.95 \\
        DDO~168 & 24.1 & 495.0 & 67.8 & - & - & ISO & 39.81 & 2.81 \\
        DDO~216 & 6.5 & 721.5 & 98.8 & - & - & ISO & 127.02 & 0.15 \\
        NGC~2366 & 15.1 & 1964.3 & 269.1 & - & - & ISO & 43.89 & 1.21 \\
        WLM & 6.2 & 1363.9 & 186.8 & - & - & ISO & 57.46 & 0.74 \\
        \midrule
        AGC~242019 & 1.9 & 4951.4 & 678.3 & - & - & NFW & 1.46 & 45.10\\
        \bottomrule
     \end{tabular}
     \begin{tablenotes}
      \small
      \item Note. Col.(2): central surface density of stellar disk; Col. (3): scale length of stellar disk; Col.(4): scale height of stellar disk; Col.(5): central volume density of bulge; Col.(6): scale radius of bulge; Col.(7): type of dark matter profile; Col.(8): c is the "concentration" of NFW profile and $\rho_0$ is the core density of ISO profile; Col.(9): $V_{\rm 200}$ is the circular velocity at $r_{200}$ of NFW profile and $r_{r_{\rm c}}$ is core radius of ISO profile. Col.(7)-(9) are from \cite{deBlok2008}, \cite{Oh2015} and \cite{Shi2021}.
    \end{tablenotes}
  \end{threeparttable}
\end{table*}

\subsection{The profile of the dark matter halo}

 The dark matter model is in general described with pseudo-isothermal (ISO) \citep{Begeman1991} or Navarro–Frenk–White (NFW) \citep{Navarro1997} profiles. 
 For spiral galaxies, we use the model with smaller reduced $\chi^{2}$ provided by \cite{deBlok2008}. But we find that the derived
 scale height is almost the same if adopting the alternative model.
  For dwarf galaxies, the ISO profiles are in general better than the NFW models \citep{Oh2011b}, so we adopt the former as provided by \cite{Oh2015}. For the UDG AGC 242019, the NFW model gives a much better fit  as compared to the ISO, and we thus adopt the NFW model.

The profile of the ISO model is
\begin{equation}
    \rho_{\rm ISO}(r) = \rho_{\rm DM,0}(1+\frac{r^2}{r^2_{\rm c}})^{-1},
    \label{eq.14}
\end{equation}
where $\rho_{\rm 0}$ is the characteristic core density and $r_{\rm c}$ is the core radius of the DM halo. 

The NFW mass distribution profile is
\begin{equation}                                       
    \rho_{\rm NFW}(r) = \frac{\rho_{\rm c}\delta_{\rm char}}{(r/r_{\rm s})+(1+r/r_{\rm s})^2},
    \label{eq.15}
\end{equation}
where $\rho_{\rm c}=3H^{2}/(8\pi G)$ is the present critical density, $\delta_{\rm char} = \frac{200c^2}{3[ln(1+c)-c/(1+c)]}$ is the dimensionless density contrast and $r_{\rm s}$ is the scale radius. $r_{\rm s}=r_{\rm 200}/c$, where $c$ is the concentration of NFW profile and $r_{\rm 200}$ is the radius of a sphere within which the mean halo density is 200$\rho_{\rm c}$. $V_{\rm 200} = (r_{\rm 200}/h^{-1} \rm kpc)\ km/s$ is the circular velocity at $r_{\rm 200}$. 

In both profiles of DM halo, the spherical radius $r$ is equal to $\sqrt{R^2+z^2}$ at given $R$ and $z$ in cylindrical coordinates.

\subsection{Numerical calculation of the scale height}\label{sec.scaleheight}

With the given model of the stellar component, we can reduce Eq.~\ref{eq.11} into two second-order ordinary differential equations in $\rho_{\rm HI}$ and $\rho_{\rm H2}$. For dwarf galaxies and AGC~242019, we only need to solve the ordinary differential equation in $\rho_{\rm HI}$.
To solve these two ordinary differential equations, we need to give initial conditions, which is a guess of $\rho_{i,0}$ at the first iteration:
\begin{equation}
    \rho_{i,0} = \frac{\Sigma_{i}}{h_{i}}, \ \frac{d\rho_{i,0}}{dz} = 0.
    \label{eq.16}
\end{equation}

With the initial guess of scale heights, we can solve the second-order ordinary differential equations of both atomic and molecular gas, and obtain the vertical volume density distribution of each component. Then, we fit the resulting vertical distribution with Gaussian function ($\rho(z)=\rho_{i,0}{\rm exp}(-\frac{z^{2}}{2h_{i}^{2}})$) to derive the new scale height. With a new scale height, we repeat the previous steps until the integration of vertical Gaussian distribution ($\int \rho(z) \ dz = \sqrt{2\pi}\rho_{i,0}h_{i}$) is within one percent difference of the observed surface densities. By running these iterations, we obtain the radial profiles of the scale height for atomic gas and molecular gas. 

For the scale height of SFRs, we assume it as the mean of the scale heights of atomic and molecular gas components weighted by their gas fractions, following the equation in \cite{Bacchini2019} as:
\begin{equation}
    h_{\rm SFR}(r) = h_{\rm HI}(r)\frac{\Sigma_{\rm HI}}{\Sigma_{\rm HI}+\Sigma_{\rm H_2}}+h_{\rm H_2}(r)\frac{\Sigma_{\rm H_2}}{\Sigma_{\rm HI}+\Sigma_{\rm H_2}}
    \label{eq.17}
\end{equation}
at a given radius. This definition for $h_{\rm SFR}$ ideally describes the flaring SFR vertical distribution of the MW \citep{Bacchini2019a}. The advantage of this equation is that it can be applied to both molecular gas dominated and atomic gas dominated regions. Figure~\ref{fig.1} shows the result of the scale heights of atomic gas, molecular gas, and SFR.

\begin{figure*}
  \centering
  \includegraphics[width=0.90\textwidth]{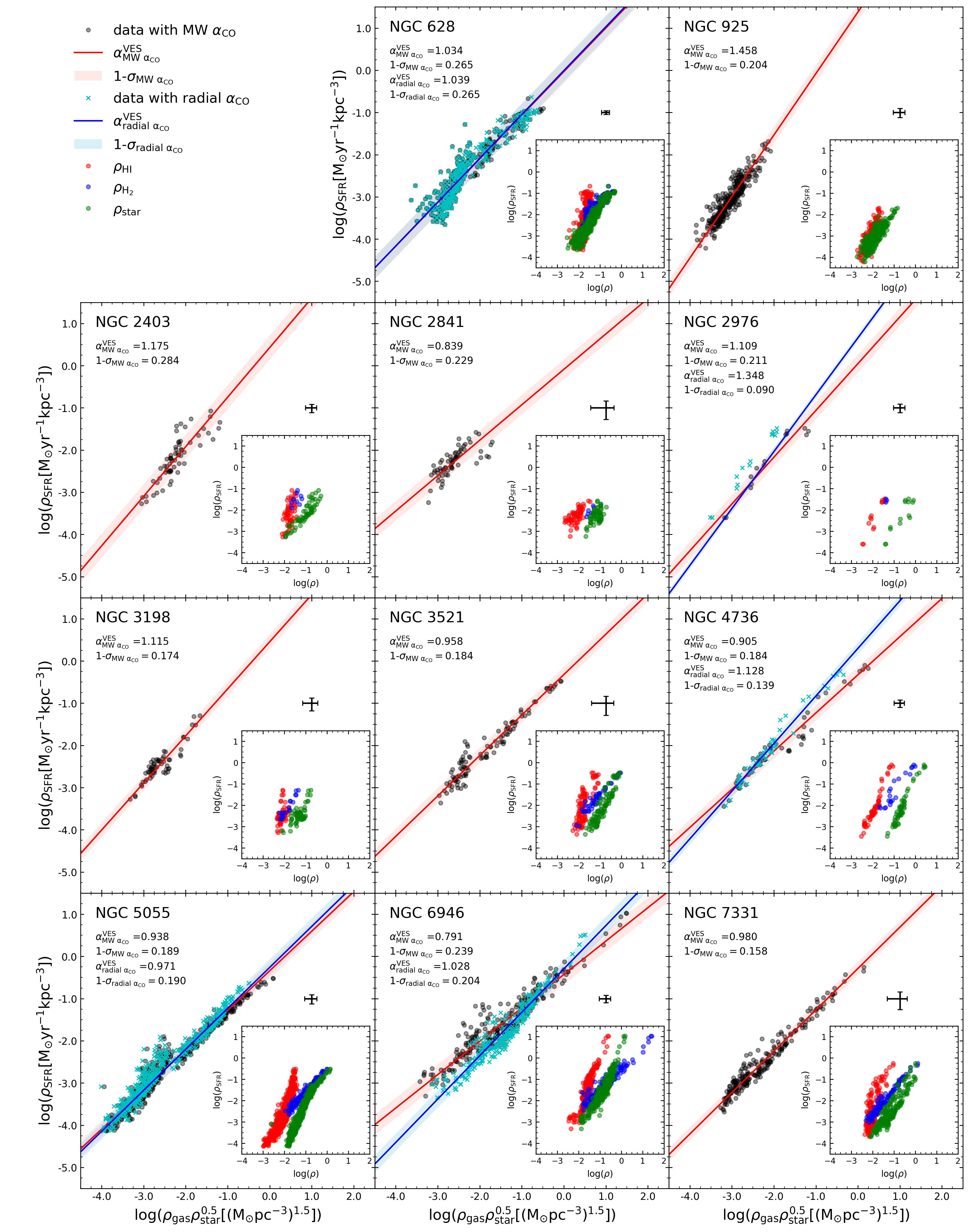}
  \caption{The volumetric ES law are plotted for every spiral galaxy in our sample. In each panel, the black points are the data with $\alpha_{\rm CO}^{1-0}$ = 4.35 M$_{\odot}$pc$^{-2}({\rm K\ km\ s^{-1}})^{-1}$, the red lines are the best-fitted models for black points and light-red regions show rms scatter bounds for each best-fitted model, $\alpha^{\rm VES}_{\rm MW\ \alpha_{CO}}$ is the slope of the best-fitted model. The subgraphs embedded in the diagram of each galaxy show the correlation between the volume densities of atomic gas, molecular gas, star and the SFR. For NGC~628, NGC~2976, NGC~4736, NGC~5055 and NGC~6946, the cyan 'x' markers are the data point with radial-dependent $\alpha_{\rm CO}^{1-0}$ from \citet{Sandstrom2013}, and the blue lines are the corresponding best-fitted models and light-blue regions show rms scatter bounds for each best-fitted model, $\alpha^{\rm VES}_{\rm radial\ \alpha_{CO}}$ is the corresponding slope. The typical errorbars are also plotted in each panel.}
  \label{fig.2}
\end{figure*}

\begin{figure*}
  \centering
  \includegraphics[width=\textwidth,height=14.0cm]{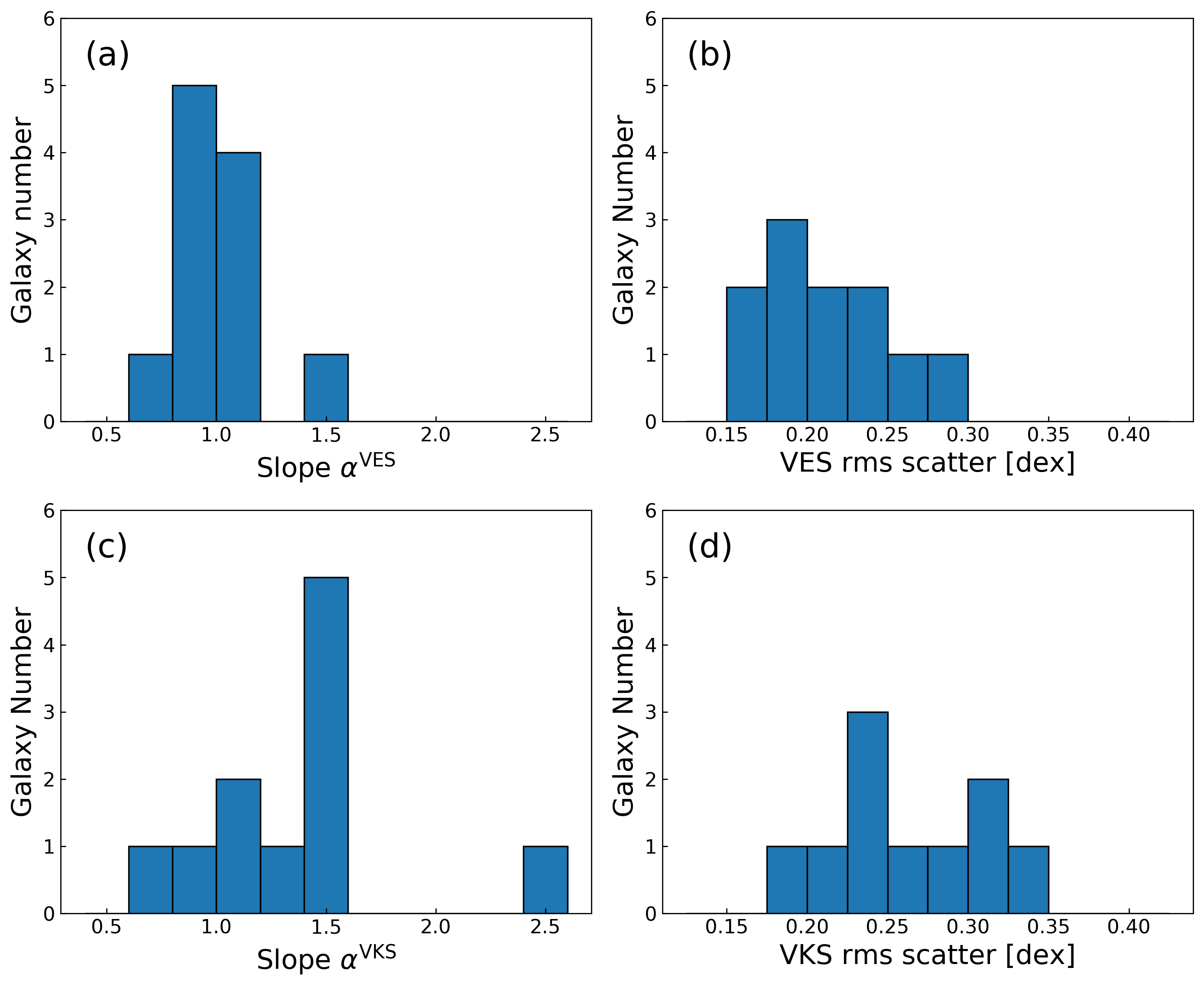}
    \caption{Panel (a) and (b) present the distribution of the slope $\alpha^{\rm VES}$ and rms scatter of the volumetric ES law at given x-axis for 11 spiral galaxies, respectively. Panel (c) and (d) present the distribution of the slope $\alpha^{\rm VKS}$ and rms scatter of the volumetric KS law at given x-axis, respectively.}
    \label{fig.3}
\end{figure*}

\section{Results}\label{sec.result}

\begin{figure*}
  \centering
  \subfloat{\includegraphics[width=0.475\textwidth]{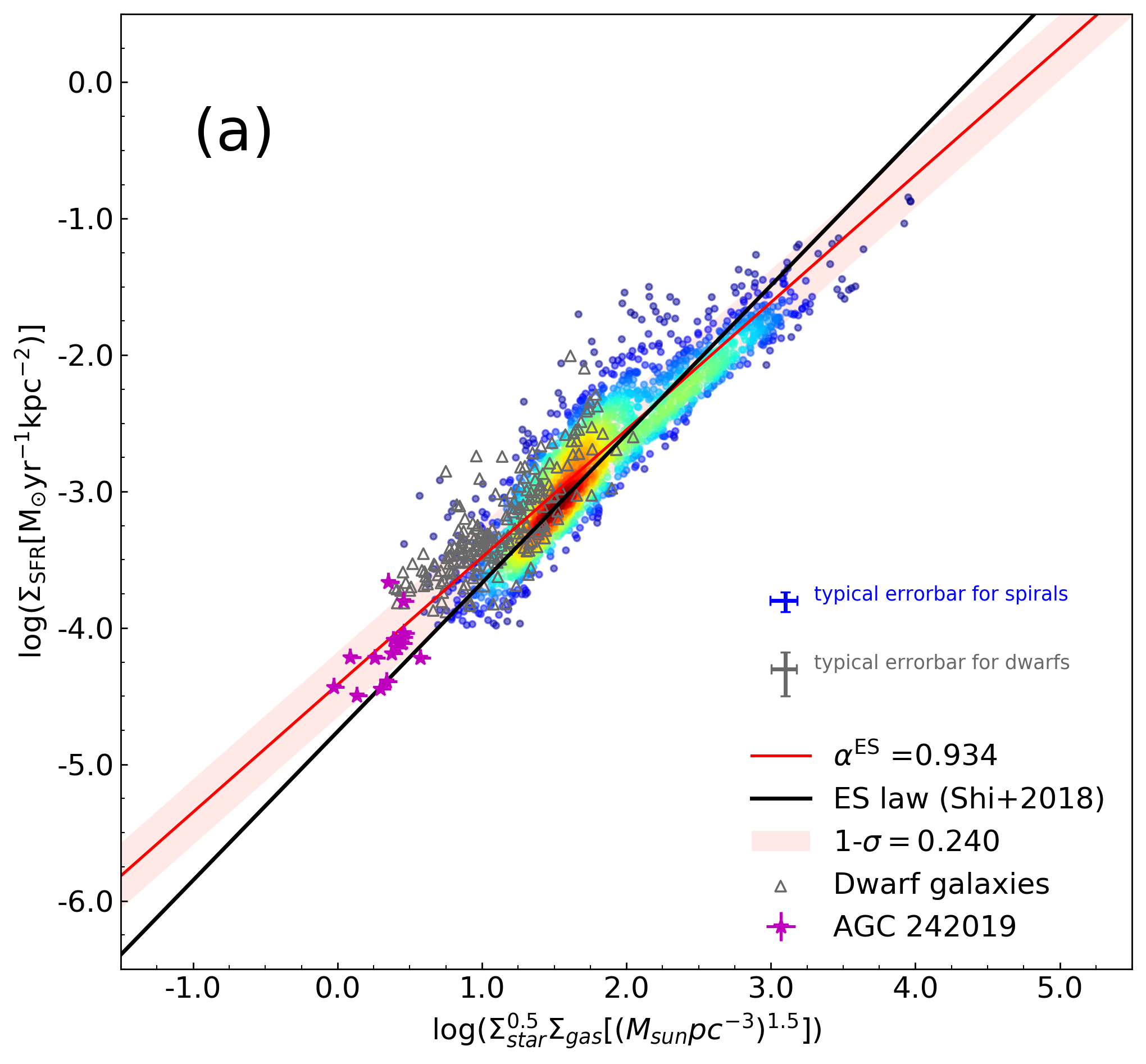}}\quad
  \subfloat{\includegraphics[width=0.475\textwidth]{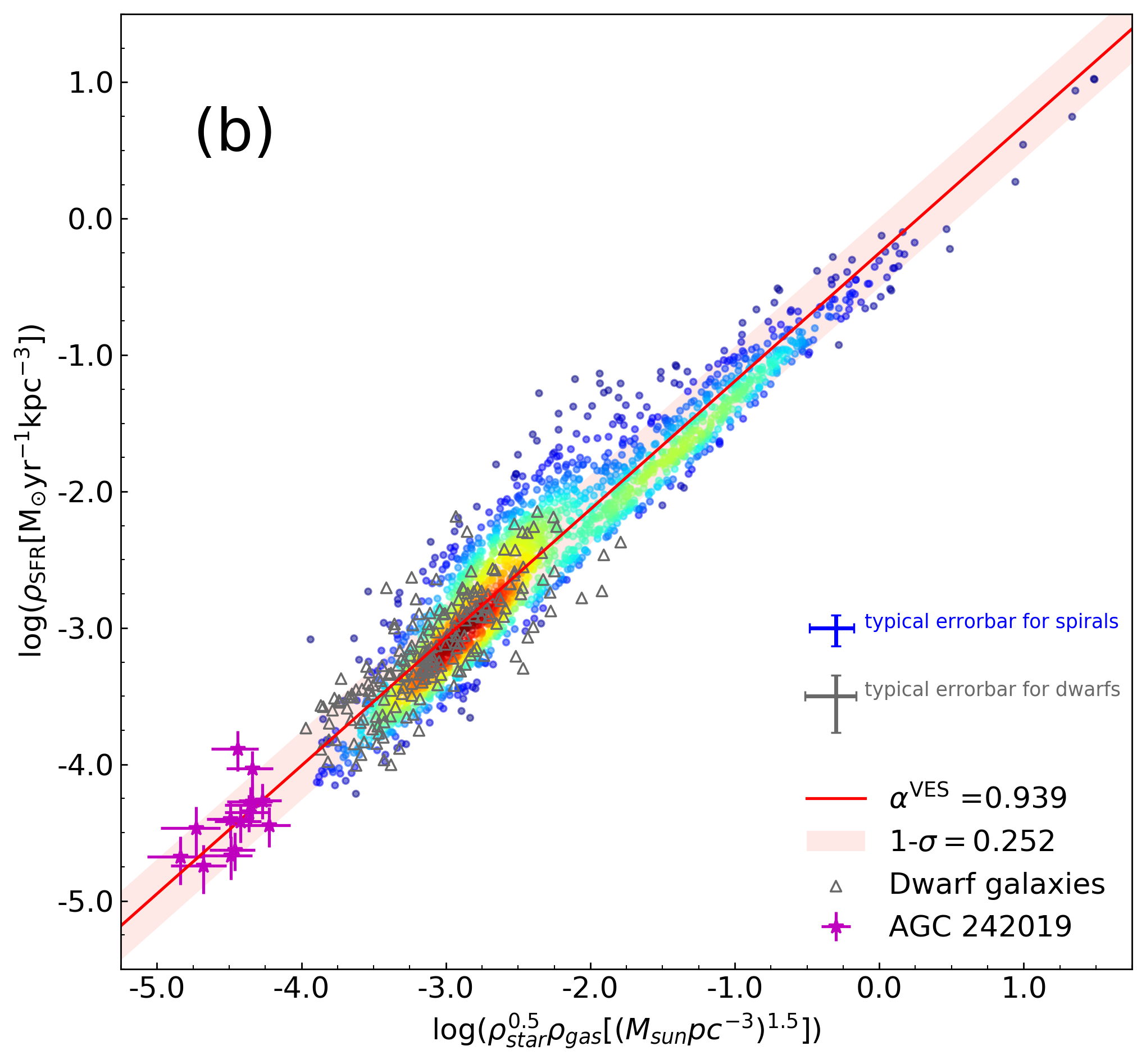}}\\
  \subfloat{\includegraphics[width=0.475\textwidth]{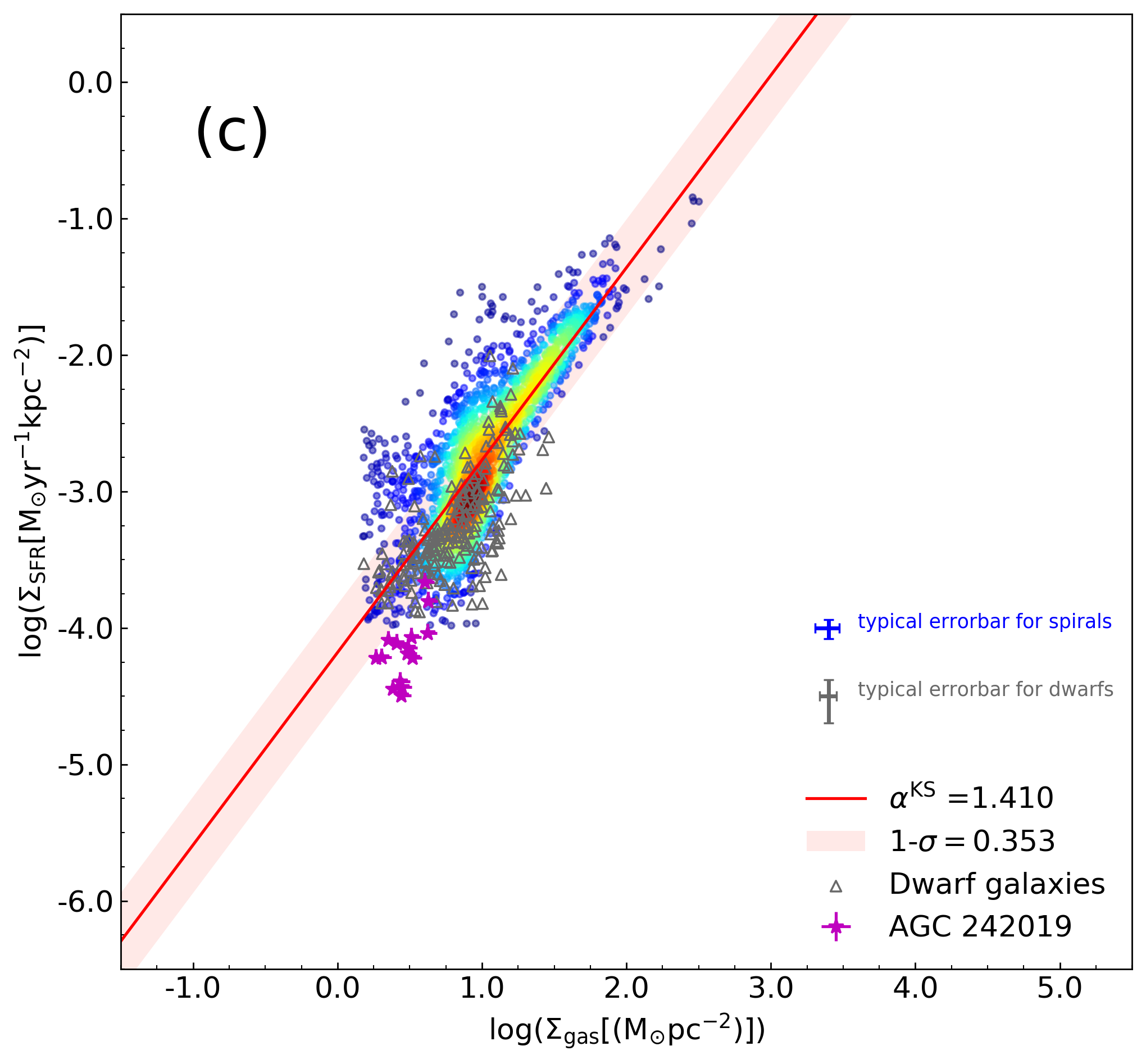}}\quad
  \subfloat{\includegraphics[width=0.475\textwidth]{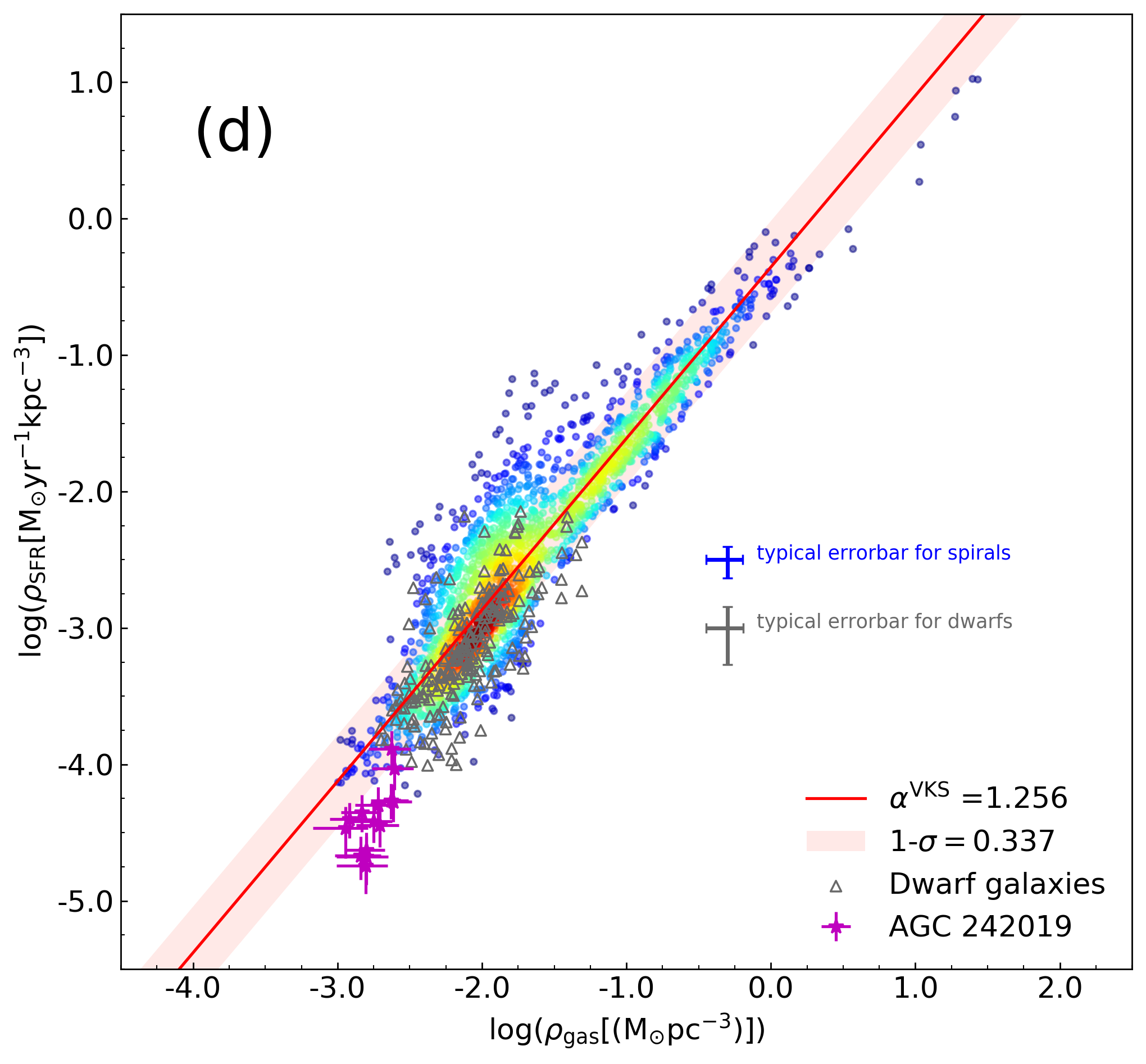}}
  \caption{Panel (a) and (b) show the ES law in surface densities and volume densities, respectively. Panel (c) and (d) show the KS law in surface densities and volume densities, respectively. Spiral galaxies are shown by dots and the color background shows the number density of the data points. Grey triangles represent dwarf galaxies and magenta stars with errorbars represent AGC~242019. In each panel, the typical errorbars of spiral galaxies and dwarf galaxies are plotted in blue and grey errorbars, respectively. The red line represents the best-fit model for our sample and the light-red region shows rms scatter bounds for the best-fit model. The black line in panel (a) is the best-fit model of surface-density ES law in \citet{Shi2018}.}
  \label{fig.4}
\end{figure*}

\begin{figure*}
  \centering
  \includegraphics[width=\textwidth,height=12.5cm]{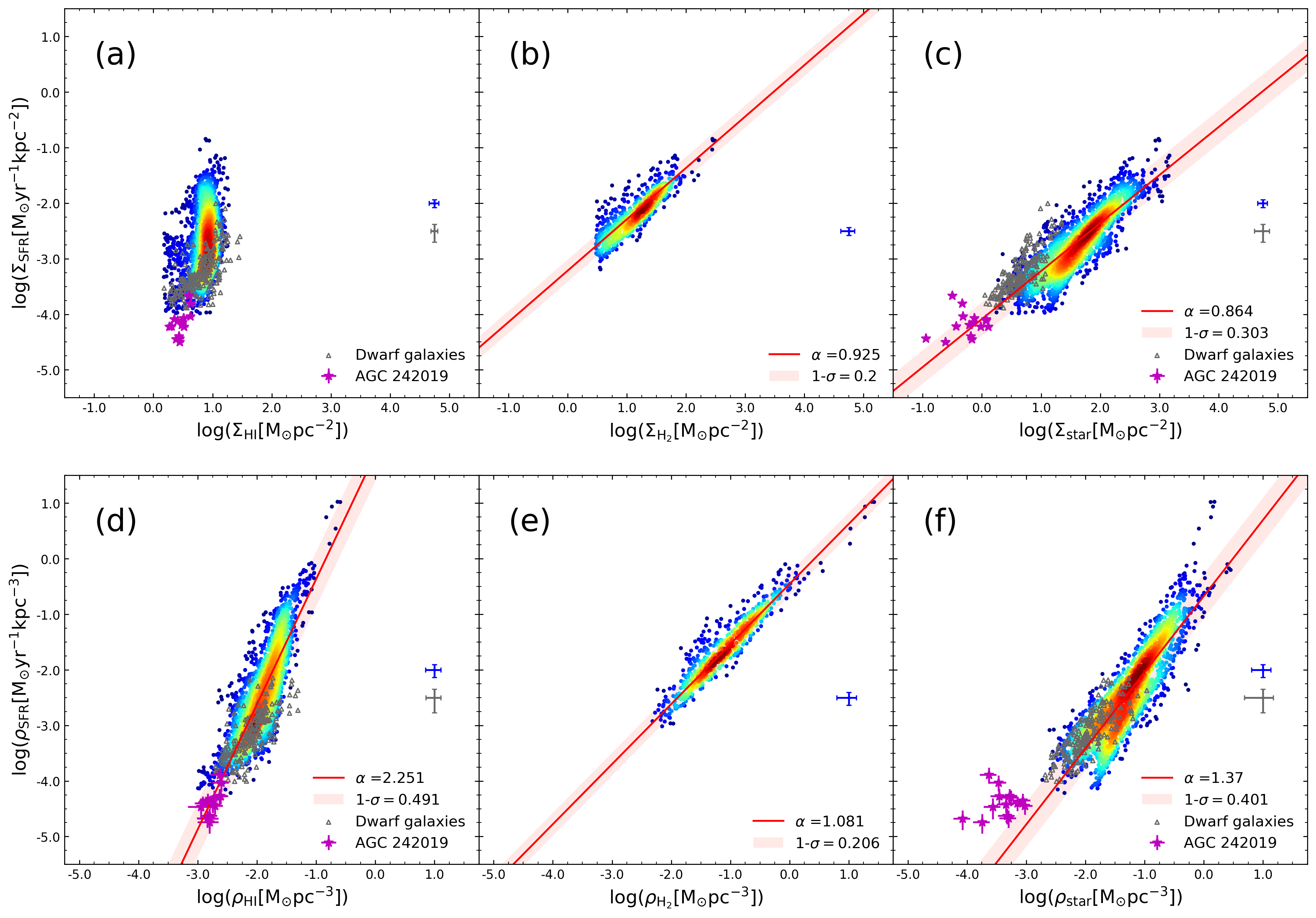}
    \caption{(a): $\Sigma_{\rm SFR}-\Sigma_{\rm HI}$; (b): $\Sigma_{\rm SFR}-\Sigma_{\rm H_2}$; (c): $\Sigma_{\rm SFR}-\Sigma_{\rm star}$; (d): $\rho_{\rm SFR}-\rho_{\rm HI}$; (e): $\rho_{\rm SFR}-\rho_{\rm H_2}$; (f): $\rho_{\rm SFR}-\rho_{\rm star}$. Spiral galaxies are shown by dots and the color background shows the number density of the data points. The grey triangles represent dwarf galaxies and magenta stars with errorbars represent AGC~242019. In each panel, the typical errorbars of spiral galaxies and dwarf galaxies are plotted in blue and grey errorbars, respectively. The red lines represent the best-fit models for each relation and the light-red region shows rms scatter bounds for the best-fit models.}
    \label{fig.5}
\end{figure*}

With the radial profile of the scale height of each component, we convert the surface densities to the volume densities. We first plot the volumetric ES law for individual spiral galaxies in Figure~\ref{fig.2} and fit them with $\rho_{\rm SFR}$ $\propto$ $(\rho_{\rm gas}\rho_{\rm star}^{0.5})^{\alpha_{\rm VES}}$. Note that
dwarf galaxies have small sizes and do not offer enough data points for us to examine the star formation law for individual objects.
Each panel in Figure~\ref{fig.2} also contains sub-graphs to show the correlations of the volume densities of atomic gas, molecular gas and stars with the volume densities of SFRs, respectively. In all spiral galaxies $\rho_{\rm SFR}$ is found to be tightly correlated with $\rho_{\rm gas}\rho_{\rm star}^{0.5}$ as shown by the best-fitted red lines. Figure~\ref{fig.3} (a) shows the distribution of the slopes ${\alpha^{\rm VES}}$, which ranges from 0.79 to 1.46, with the median and standard deviation of  0.98 and 0.18, respectively. There are two outlier, i.e., NGC~6946 with ${\alpha^{\rm VES}}$ $\sim$ 0.79$\pm$0.02 and NGC~925 with ${\alpha^{\rm VES}}$ $\sim$ 1.46$\pm$0.04. As shown in the figure, the adopted average $\alpha_{\rm CO}^{1-0}$ may lead to overestimation and underestimation of molecular gas masses for in the inner and outer regions of NGC~6946, respectively. Across the disk of NGC~925, the atomic gas dominates and molecular gas is very faint. Figure~\ref{fig.3} (b) shows the distribution of the rms scatter of the relationship around the best fit at given x-axis for individual spiral galaxies. It ranges from  0.15 dex to 0.28 dex, with a median of 0.20 dex. As a comparison, we also carry out the volumetric KS law fitting with a form of $\rho_{\rm SFR}$ $\propto$ $(\rho_{\rm gas})^{\alpha^{\rm VKS}}$ for individual spiral galaxies. As shown in Figure~\ref{fig.3} (c), ${\alpha^{\rm VKS}}$ ranges from 0.72 to 2.41 and has a median and standard deviation of 1.42 and 0.41, respectively. The latter is more than a factor of two larger than the standard deviation of the volumetric ES law slopes $\alpha^{VES}$. As shown in Figure~\ref{fig.3} (d), the  rms scatter of the volumetric KS law around the best fit at given x-axis ranges from 0.18 dex to 0.43 dex with a median of 0.27 dex, that is a factor of 1.35 larger than the median of the volumetric ES law rms scatters.

We also investigate the surface-density and volumetric ES law for spatially-resolved regions in all 25 galaxies including spirals, dwarfs and one UDG. As shown in the panel (a) and (b) of Figure~\ref{fig.4}, the rms scatter of the surface-density ES law around the best fit at given x-axis is slightly smaller than that of the volumetric ES law, with values of $0.24$ dex and $0.25$ dex, respectively. Slopes of both surface-density and volumetric ES law are 0.93 and 0.94, all close to 1.

We also plot the KS law in surface densities and volume densities for the whole sample in Figure \ref{fig.4}. In general, there are also good correlations between the densities of the SFR and total gas mass, but low-density points show relatively large deviations in both surface-density and volumetric KS law, especially the diffuse galaxy AGC~242019. The slopes of the KS law in surface densities and volume densities are $\sim 1.41$ and $\sim 1.26$, with rms scatter around the best fit at given x-axis are 0.35 and 0.34 dex. Compared to the volumetric KS law, the volumetric ES law has a slightly smaller scatter. \cite{Bacchini2019} has studied the volumetric KS law for galaxies in our sample and reported orthogonal intrinsic scatter $1-\sigma_{\perp}$ of 0.12 dex ( $1-\sigma_{\perp} \sim$ 0.18 dex in our work). \citet{Bacchini2019} performs a radial analysis using azimuthally-averaged quantities, while we use the local measurements for surface quantities. We also perform the radial analysis in the Appendix and Figure~\ref{fig.A1} shows similar results with Figure~\ref{fig.4}. There is some differences in the method to derive the scale heights: \cite{Bacchini2019} fitted the surface densities and velocity dispersion radial profiles with given functions, leading to smooth radial profiles of scale heights; while we directly use the measurements of surface densities and the \texttt{$^{\rm 3D}$Barolo} fitting results of velocity dispersion. This may contribute to the difference between our results and \cite{Bacchini2019}. The slope of the volumetric KS law in our work is $1.26$, which is smaller than the slope ($1.34-1.91$) in \citet{Bacchini2019} but similar to the slope ($1.26$) for five edge-on galaxies in \citet{Yim2020}. 

Figure~\ref{fig.5} also shows the relationships of HI-SFR, H$_2$-SFR and star-SFR in both surface densities and volume densities. Although the SFR generally increases with the atomic gas mass on global scale \citep{Zhou2018}, no or little dependence of the SFR surface density on atomic gas surface density is found on sub-kiloparsec scale \citep{Bigiel2008}, as also shown in panel (a) and (d) of Figure~\ref{fig.5}. However, as already found in \citet{Bacchini2019,Bacchini2019a}, we confirm a correlation between $\rho_{\rm HI}$ and $\rho_{\rm SFR}$. Due to the differences between our methods to derive the volume density, our slope of 2.26 for $\rho_{\rm HI}-\rho_{\rm SFR}$ is different from the slope of 2.8 in \citet{Bacchini2019}. In the panel (b) and (e) of Figure~\ref{fig.4}, the slope of $\rho_{\rm H_2}-\rho_{\rm SFR}$ is found to be 1.08, also different from 0.73 in \citet{Bacchini2019}. In the panel (c) and (f) of Figure~\ref{fig.5}, we show that there are also relationships between the SFR and the stellar mass in both surface densities and volume densities.

\section{Discussion}\label{sec.discusssion}

\subsection{Factors that influence the result}

\subsubsection{Different measurements of the SFR scale heights }\label{sec.diff_hSFR}

Molecular gas is the pivotal fuel of star formation. Several studies \citep{Gao2004,Gao2004a} suggested that the SFR is tightly related to the dense molecular gas. Besides, \citet{Bigiel2008} and \citet{Leroy2008} found a good correlation between $\Sigma_{\rm H_2}$ and $\Sigma_{\rm SFR}$ in nearby spiral galaxies, i.e., the so-called molecular Schmidt law, indicating that the star formation process is closely connected with the content of molecular gas, and the scale height of SFR should be close to that of molecular gas. As a result we further compare our fiducial result with a case where $h_{\rm SFR}=h_{\rm H_2}$ for spiral galaxies as listed in Table~\ref{tab.3}. Compared to the fiducial case (Eq.~\ref{eq.17}) where $\alpha^{\rm VES}$=0.90$\pm$0.01, the case with $h_{\rm SFR}=h_{\rm H_2}$ shows a slightly smaller slope with $\alpha^{\rm VES}$ of 0.83$\pm$0.01. In the galaxy periphery, the atomic gas becomes dominant and the SFR scale height in the fiducial case is much larger than the scale height of molecular gas, thus resulting in a smaller slope when assuming $h_{\rm SFR}=h_{\rm H_2}$. The rms scatters of the volumetric ES law under these two different $h_{\rm SFR}$ cases are more or less the same. The scale heights of the SFR thus have no significant influence on the overall result.

\subsubsection{Flaring stellar disks}\label{sec.diff_hstar}

Some previous studies of edge-on galaxies found that the scale height of stellar disks does not vary with radius \citep{deGrijs1997,Kregel2002}. However, \citet{Yim2011,Yim2014,Yim2020} found somewhat flaring stellar disks for five edge-on galaxies with high-spatial resolved observations. In our work, we also assume that the scale heights of stellar disks are constant. Although the different assumptions of stellar scale height hardly lead to remarkable changes on the scale height of atomic gas and molecular gas, they do affect the stellar mass volume density on the mid-plane. The average gradient of flaring stellar disks is about $\frac{30\ {\rm pc}}{\rm kpc}$ for the four edge-on spiral galaxies in \citet{Yim2014}. We present the volumetric ES law with flaring stellar disks in the form of $h(R)=\frac{30\ {\rm pc}}{\rm kpc}\times (R-l_{\rm star})+l_{\rm star}/7.3$ as shown in Table~\ref{tab.3}. Compared to the result with constant stellar disks, the slope and rms scatter of the volumetric ES law with flaring disks have no remarkable difference. 

\subsubsection{Radial-dependent CO-to-H$_2$ conversion factor}\label{sec.diff_alphaCO}

\begin{figure}
  \centering
  \includegraphics[width=0.9\linewidth]{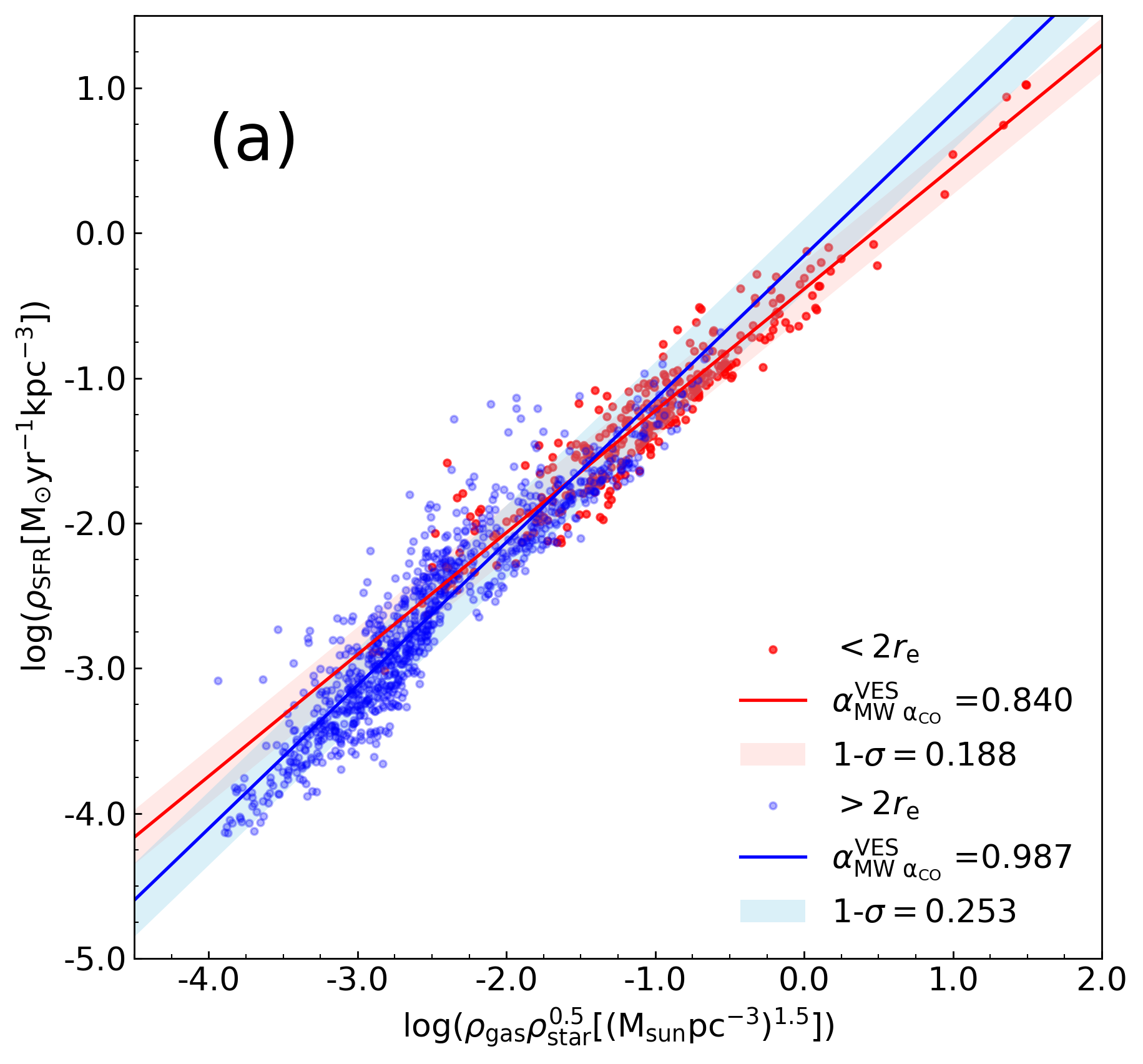}\hfill
  \includegraphics[width=0.9\linewidth]{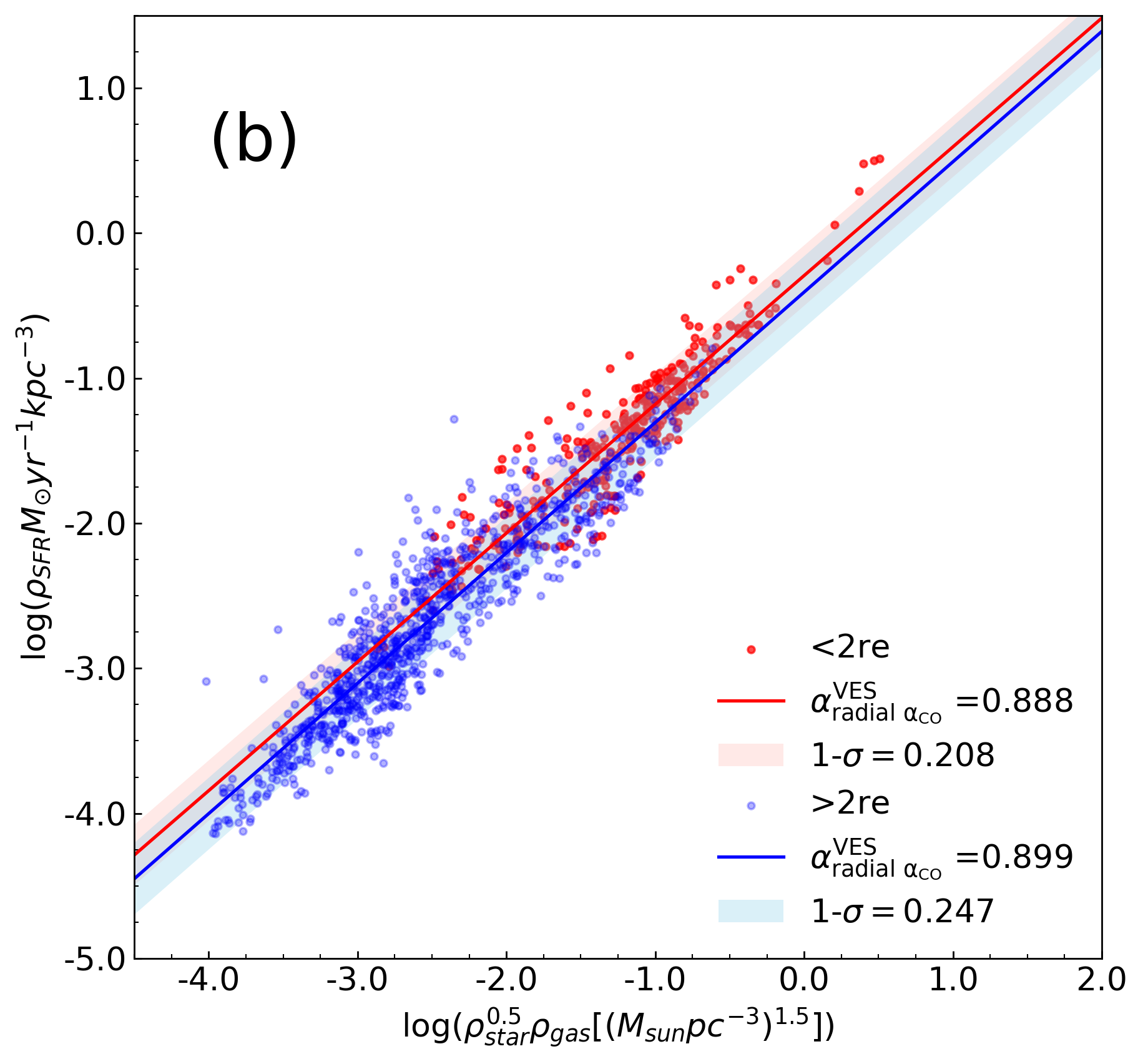}
\caption{The volumetric ES law with different CO-to-H$_2$ conversion factors for NGC~628, NGC~3521, NGC~4736, NGC~5055 and NGC~6946. Panel (a) shows the volumetric ES law for regions $<2r_{\rm e}$ and $>2r_{\rm e}$ with average galactic conversion factor ($\alpha_{\rm CO}^{1-0}$ =4.35 M$_{\odot}$pc$^{-2}({\rm K\ km\ s^{-1}})^{-1}$). Panel (b) shows the volumetric ES law for regions $<2r_{\rm e}$ and $>2r_{\rm e}$ with $\alpha_{\rm CO}$ from \citet{Sandstrom2013}, respectively.}
  \label{fig.6}
\end{figure}

For NGC~628, NGC~3521, NGC~4736, NGC~5055 and NGC~6946, the volumetric ES law seems to change slope when moving from the inner to the outer regions of their disks. This phenomenon also exists in the surface-density ES law. In the subgraphs of each panel of Figure~\ref{fig.2}, we show the relationships of $\rho_{\rm HI}$, $\rho_{\rm H_2}$, $\rho_{\rm star}^{0.5}$ with $\rho_{\rm SFR}$ separately. For galaxies with clear inflection in their ES law, it is found that the different slopes are due to the inflection trend of total gas: in the inner regions molecular gas dominates and in outer regions atomic gas dominates; this combined with the fact that the slopes of $\rho_{\rm HI}$ - $\rho_{\rm SFR}$ and $\rho_{\rm H_2}$ - $\rho_{\rm SFR}$ are different, leads to the inflection in the ES law. Such an inflection might be due to the constant CO-to-H$_2$ conversion factor $\alpha_{\rm CO}$  that we adopt, which does not consider the metallicity dependence (e.g. \citet{Shi2016}). It is found that $\alpha_{\rm CO}$ appears to drop in the central regions of spiral galaxies \citep{Sandstrom2013,Bolatto2013}. As a result, we adopt the radial profile of $\alpha_{\rm CO}$ based on spectral line modeling and dust observations in \citet{Sandstrom2013} and plot the volumetric ES law for NGC~628, NGC~2976, NGC~4736, NGC~5055, and NGC~6946 in Figure~\ref{fig.2} with blue plus markers and blue lines. It is shown that the radial-dependent $\alpha_{\rm CO}$ alleviates the inflection trend and increases the overall slope.
 We further present the volumetric ES law for these five galaxies as a whole with average $\alpha_{\rm CO}$ of the MW and radial-dependent $\alpha_{\rm CO}$ separately in Figure~\ref{fig.6}. We divide the data points into two parts: $r<2r_{\rm e}$ and $r>2r_{\rm e}$. For the result with average $\alpha_{\rm CO}$ of the MW, the outer regions ($r>2r_{\rm e}$) have a larger slope than inner regions ($r<2r_{\rm e}$). However, for the result with radial-dependent $\alpha_{\rm CO}$, the slopes for outer and inner regions are similar, indicating that the inflection of the correlation is likely due to the over-estimation of the molecular gas mass in the central regions of galaxies when adopting constant $\alpha_{\rm CO}$. The slopes of best-fit model for all the data points in panel (a) and (b) of Figure~\ref{fig.6} are 0.945 and 0.940, respectively, as listed in Table~\ref{tab.3}.

\subsection{Physical Implications}\label{sec.theory}

\begin{figure}
  \centering
  \includegraphics[width=0.9\linewidth]{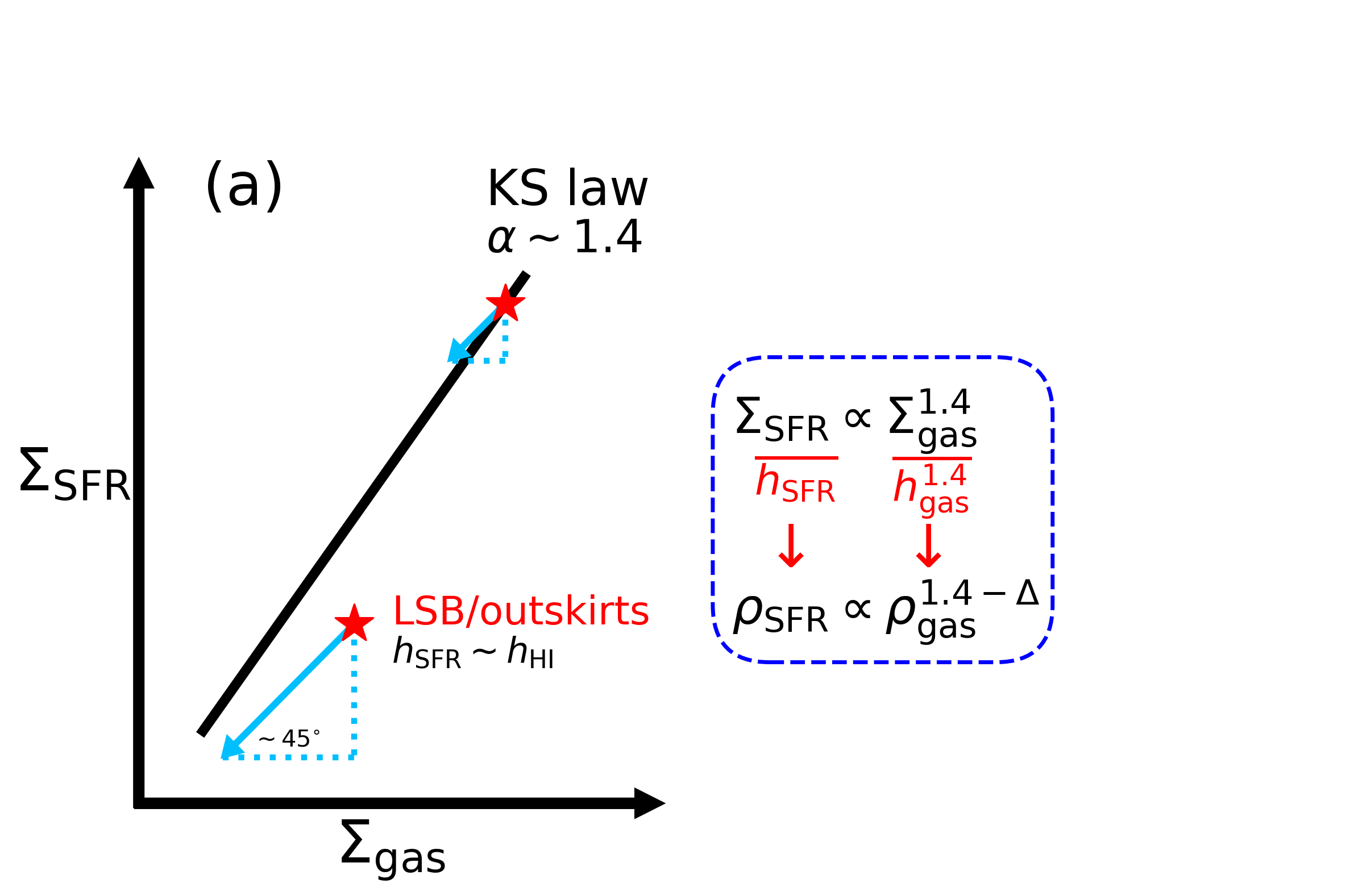}
  \includegraphics[width=0.9\linewidth]{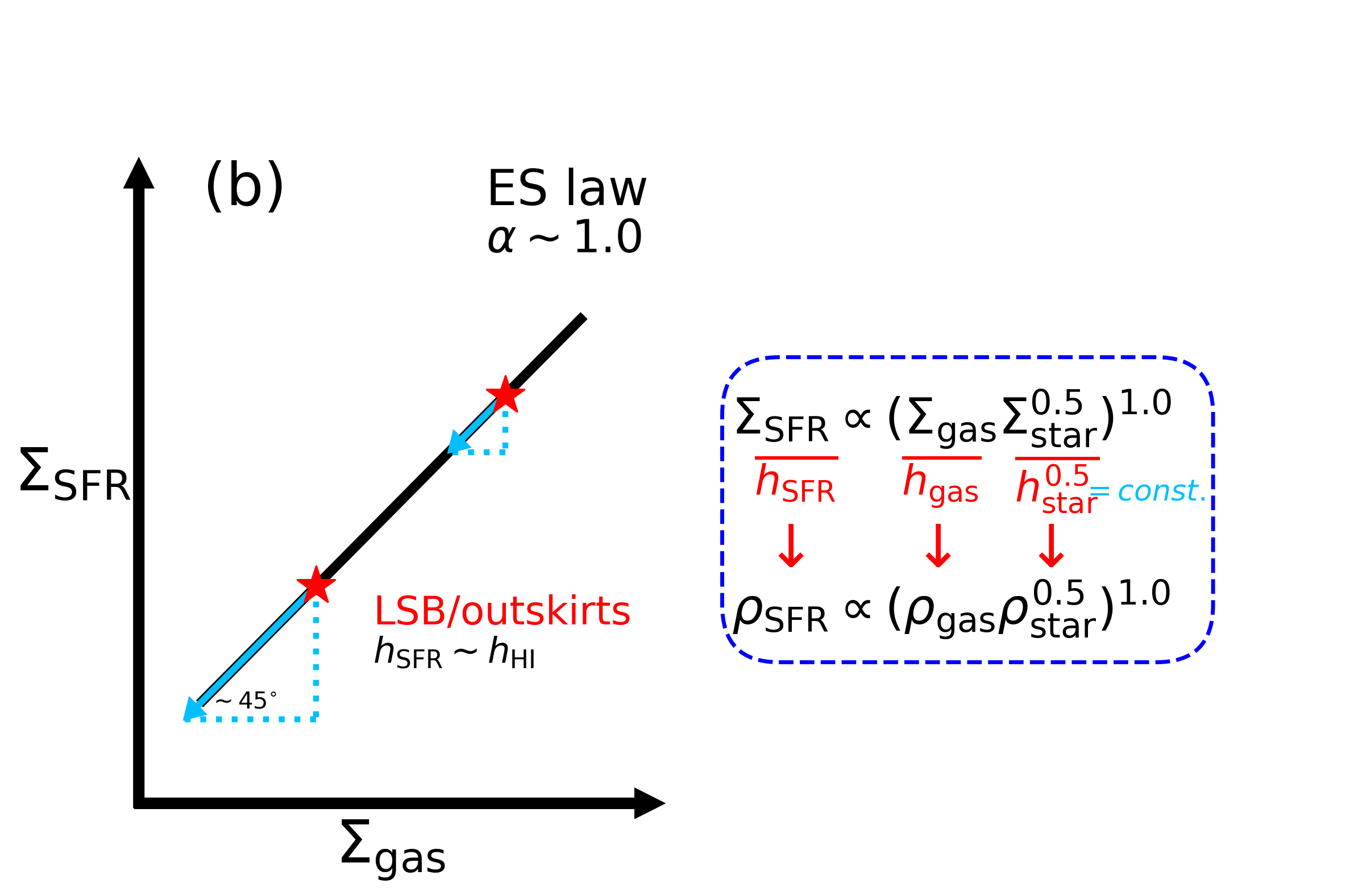}
  \caption{Panel (a) shows shift of data points on the KS law from surface densities to volume densities. Panel (b) shows the shift of data points on the ES law from surface densities to volume densities. }
  \label{fig.7}
\end{figure}

As found in many works \citep{Kennicutt1998,Wyder2009,Onodera2010,Shi2018}, the KS law in surface densities breaks down at the low-density end. \citet{Bacchini2020} claims that the KS law holds unbroken at low-density end when considering the volume densities. This indicates that the break in the surface-density KS law is due to the disk flaring rather than the decrease of star formation efficiency at the low-density end, which was also suggested by \citet{Elmegreen2015}. As shown in Figure~\ref{fig.5}, the breakdowns seem to appear both on surface-density and volumetric KS law. The median offsets of dwarf galaxies from the surface-density and volumetric KS law are moderate, i.e., 0.30 dex and 0.18 dex, respectively. For AGC~242019, the median offsets from surface-density and volumetric KS law reaches as high as 0.64 dex and 0.59 dex, respectively. 
The median offsets of these objects in the KS law reduce when we convert the surface densities to volume densities, indicating that the projection effects do contribute to the breakdown of surface-density KS law at low-density end to some level. However, based on our results, the deviation of low-density regions from KS law is not entirely caused by the projection effect. 
Most of the large deviations occur at regions with larger $\rho_{\rm gas}$ (compared to $\rho_{\rm SFR}-\rho_{\rm gas}$) but much smaller $\rho_{\rm star}$ (compared to $\rho_{\rm SFR}-\rho_{\rm star}$). 
Unlike the KS law, surface-density and volumetric ES law holds unbroken in dwarf galaxies and even AGC~242019, implying that the stellar mass density plays an significant role in regulating the star formation process in low-density environments.

As shown in panel (a) of Figure~\ref{fig.7}, when we convert surface densities to volume densities, points at different surface densities move by a different amount. The low-surface-brightness and outskirt point moves significantly at an about 45-degree angle to the lower left and come closer to the relationship in the volume density. And the KS law is transformed from $\Sigma_{\rm SFR}\propto\Sigma_{\rm gas}^{1.4}$ to $\rho_{\rm SFR}\propto\rho_{\rm gas}^{1.4-\Delta}$ after dividing by the scale height where $\Delta$ $>$ 0, resulting in a decrease of the slope. For the ES law, since the slope in surface densities is close to unity and the $h_{\rm star}$ is constant, the slope of the ES law remains unity from surface density to volume density but the scatter increases because of the observational error in $h_{\rm star}$.

The star formation process is the gravitational collapse of gas, which can be written as $\rho_{\rm SFR}=\frac{\rho_{\rm gas}}{\tau}$ ($\tau$ is the star formation timescale). The unity
slope of the volumetric ES law indicates that $\tau \propto \frac{1}{\rho_{\rm star}^{0.5}}$,
indicating that the star formation efficiency ($\equiv$ $\rho_{\rm SFR}/\rho_{\rm gas}$) is regulated by the stellar mass density instead of other parameters. The role of stellar mass in regulating star formation may be achieved through the stellar gravity such as in self-regulated star formation models \citep{Ostriker2010,Kim2011} as illustrated in \citet{Shi2018}.

\begin{table*}
  \begin{threeparttable}
    \caption{The parameters of best fit}
    \label{tab.3}
     \begin{tabular}{clcccc}
        \toprule
        \begin{tabular}[l]{@{}l@{}}Figures\\(1)\end{tabular} & 
        \begin{tabular}[l]{@{}l@{}}Correlations\\(2) \end{tabular}& \begin{tabular}[l]{@{}l@{}}$\beta$\\(3)\end{tabular} & \begin{tabular}[l]{@{}l@{}}$\alpha$\\(4)\end{tabular} & \begin{tabular}[l]{@{}l@{}}$\sigma$\\(5)\end{tabular} &
        \begin{tabular}[l]{@{}l@{}}r\\(6)\end{tabular} \\
        \midrule
        Fig.~\ref{fig.5}(b) & $\Sigma_{\rm SFR}-\Sigma_{\rm H_2}$ & -3.236 & 0.925 & 0.200 & 0.860\\ 
        Fig.~\ref{fig.5}(c) & $\Sigma_{\rm SFR}-\Sigma_{\rm star}$ & -4.093 & 0.840 & 0.303 & 0.859 \\ 
        Fig.~\ref{fig.4}(c) & $\Sigma_{\rm SFR}-\Sigma_{\rm gas}$ & -4.180 & 1.410 & 0.353 & 0.803 \\ 
        Fig.~\ref{fig.4}(a) & $\Sigma_{\rm SFR}-\Sigma_{\rm gas}\Sigma_{\rm star}^{0.5}$ & -4.416 & 0.934 & 0.240 & 0.914 \\ 
        Fig.~\ref{fig.A1}(c) & $\Sigma_{\rm SFR}-\Sigma_{\rm gas}$ (azimuthally-averaged) & -4.483 & 1.737 & 0.386 & 0.846 \\ 
        Fig.~\ref{fig.A1}(a) & $\Sigma_{\rm SFR}-\Sigma_{\rm gas}\Sigma_{\rm star}^{0.5}$ (azimuthally-averaged) & -4.395 & 0.972 & 0.259 & 0.933 \\
        
        \midrule
        
        Fig.~\ref{fig.5}(d) & $\rho_{\rm SFR}-\rho_{\rm HI}$ & 1.847 & 2.251 & 0.491 & 0.819 \\
        Fig.~\ref{fig.5}(e) & $\rho_{\rm SFR}-\rho_{\rm H_2}$ & -0.413 & 1.081 & 0.206 & 0.945 \\
        Fig.~\ref{fig.5}(f) & $\rho_{\rm SFR}-\rho_{\rm star}$ & -0.629 & 1.370 & 0.401 & 0.883 \\ 
        Fig.~\ref{fig.4}(d) & $\rho_{\rm SFR}-\rho_{\rm gas}$ & -0.355 & 1.256 & 0.337 & 0.918 \\ 
        Fig.~\ref{fig.4}(b) & $\rho_{\rm SFR}-\rho_{\rm gas}\rho_{\rm star}^{0.5}$ & -0.252 & 0.939 & 0.252 & 0.955 \\ 
        Fig.~\ref{fig.6}(a) & $\rho_{\rm SFR}-\rho_{\rm gas}\rho_{\rm star}^{0.5}$ (MW $\alpha_{\rm CO}$) & -0.265 & 0.945 & 0.242 & 0.964 \\ 
        Fig.~\ref{fig.6}(a) & $\rho_{\rm SFR}-\rho_{\rm gas}\rho_{\rm star}^{0.5}$ (MW $\alpha_{\rm CO}$, $r <2 r_{\rm e}$) & -0.385 & 0.840 & 0.188 & 0.953 \\
        Fig.~\ref{fig.6}(a) & $\rho_{\rm SFR}-\rho_{\rm gas}\rho_{\rm star}^{0.5}$ (MW $\alpha_{\rm CO}$, $r >2 r_{\rm e}$) & -0.156 & 0.987 & 0.253 & 0.933 \\ 
        Fig.~\ref{fig.6}(b) & $\rho_{\rm SFR}-\rho_{\rm gas}\rho_{\rm star}^{0.5}$ (radial $\alpha_{\rm CO}$) & -0.282 & 0.940 & 0.243 & 0.960 \\ 
        Fig.~\ref{fig.6}(b) & $\rho_{\rm SFR}-\rho_{\rm gas}\rho_{\rm star}^{0.5}$ (radial $\alpha_{\rm CO}$, $r <2 r_{\rm e}$) & -0.292 & 0.888 & 0.208 & 0.926 \\
        Fig.~\ref{fig.6}(b) & $\rho_{\rm SFR}-\rho_{\rm gas}\rho_{\rm star}^{0.5}$ (radial $\alpha_{\rm CO}$, $r >2 r_{\rm e}$) & -0.406 & 0.899 & 0.247 & 0.929 \\
        Sec.~\ref{sec.diff_hSFR}  & $\rho_{\rm SFR}-\rho_{\rm gas}\rho_{\rm star}^{0.5}$ ($h_{\rm SFR}=f_{\rm HI}h_{\rm HI}+f_{\rm H_2}h_{\rm H_2}$, $r < r_{\rm H_2}$) & -0.326 & 0.903 & 0.240 & 0.954 \\ 
        Sec.~\ref{sec.diff_hSFR}  & $\rho_{\rm SFR}-\rho_{\rm gas}\rho_{\rm star}^{0.5}$ ($h_{\rm SFR}=h_{\rm H_2}$, $r < r_{\rm H_2}$) & -0.243 & 0.830 & 0.234 & 0.949\\
        Sec.~\ref{sec.diff_hstar}  & $\rho_{\rm SFR}-\rho_{\rm gas}\rho_{\rm star}^{0.5}$ (flaring stellar disk) & -0.249 & 0.916 & 0.244 & 0.959 \\
        Fig.~\ref{fig.A1}(a)  & $\rho_{\rm SFR}-\rho_{\rm gas}\rho_{\rm star}^{0.5}$ (azimuthally-averaged) & -0.059 & 0.989 & 0.272 & 0.968 \\
        Fig.~\ref{fig.A1}(b)  & $\rho_{\rm SFR}-\rho_{\rm gas}$ (azimuthally-averaged) & -0.085 & 1.416 & 0.358 & 0.944 \\
        \bottomrule
     \end{tabular}
    \begin{tablenotes}
      \small
      \item Note. Col.(1): The corresponding figure or section to the correlation. Col.(2): The Y-X correlations. The correlations labeled with MW $\alpha_{\rm CO}$ and radial $\alpha_{\rm CO}$ are only for NGC~628, NGC~3521, NGC~4736, NGC~5055, and NGC~6946. The MW $\alpha_{\rm CO}$ and radial $\alpha_{\rm CO}$ mean that the conversion factor adopted in these relations are the average galactic $\alpha_{\rm CO}$ and radial-dependent $\alpha_{\rm CO}$ of these five galaxies from \citet{Sandstrom2013}, respectively. The correlations labeled with $r < r_{\rm H_2}$ only considered the data points  within the radius, of which the scale height of molecular gas is obtained in Fig.~\ref{fig.1}. Col.(3) and (4): log(Y) = $\beta$ + $\alpha$*log(X). Col.(5): The observed standard deviation from the best fit. Col.(6): The Pearson correlation coefficient.
    \end{tablenotes}
  \end{threeparttable}
\end{table*}

\section{Conclusion}\label{sec.conclusion}

We investigate the volumetric ES law for spatially resolved regions in 11 spiral galaxies, 13 dwarf galaxies and one ultra-diffuse galaxy AGC~242019. We calculate the scale height of gas disks, stellar disks, and SFR based on a gravitational coupled three-component galactic disk model under hydro-static equilibrium. The main
conclusions are:

(1) Among individual spiral galaxies, the slope of the volumetric ES law has a median of 0.98 and standard deviation of 0.18, which is smaller than the standard deviation of the volumetric KS slope ($\alpha^{\rm VKS}$=1.42$\pm$0.41).

(2) For all regions in galaxies as a whole, the rms scatter of the volumetric ES law at given x-axis is 0.25 dex, also smaller than that of the volumetric KS law (0.34 dex).
At the extreme low gas density in the UDG, the volumetric KS law breaks down but the
volumetric ES law still holds.

(3) After examining different assumptions of the SFR scale height, stellar disk scale height and the CO-to-H$_2$ conversion factors, it is found that the volumetric ES law has an unity slope and the overall scatter of the relationship at given x-axis is smaller than that of the volumetric KS law. The adoption of the radial-dependence CO-to-H$_2$ conversion factor does alleviate the inflection trend seen in the volumetric ES law of some galaxies.

(4) Compared to the surface density ES law, the volumetric ES law has a slightly larger scatter. We argue that it is due to the intrinsic unity slope of the relationship, along with the fact that the measurement error in the scale height increases the scatter of the relationship when converting to the volume density. 

(5) The unity slope of the volumetric ES law implies that the star formation efficiency is regulated by quantities related to the square root of the stellar mass density.

\section*{Acknowledge}\label{sec.acknowledge}
K.D. and Y.S. acknowledge  the support from the  National Key  R\&D Program of  China  (No.  2018YFA0404502),  the National  Natural  Science  Foundation of  China (NSFC  grants  11825302,  12141301, 12121003, 11733002),  and the science  research grants  from  the China  Manned  Space Project  with NO. CMS-CSST-2021-B02. Y.S.  thanks the  Tencent Foundation through the XPLORER PRIZE.

\section*{DATA AVAILABILITY}
The data underlying this article are available in the article.

\bibliographystyle{mnras}
\bibliography{SF_law.bib}

\begin{thebibliography}{}
\makeatletter
\relax
\def\mn@urlcharsother{\let\do\@makeother \do\$\do\&\do\#\do\^\do\_\do\%\do\~}
\def\mn@doi{\begingroup\mn@urlcharsother \@ifnextchar [ {\mn@doi@}
  {\mn@doi@[]}}
\def\mn@doi@[#1]#2{\def\@tempa{#1}\ifx\@tempa\@empty \href
  {http://dx.doi.org/#2} {doi:#2}\else \href {http://dx.doi.org/#2} {#1}\fi
  \endgroup}
\def\mn@eprint#1#2{\mn@eprint@#1:#2::\@nil}
\def\mn@eprint@arXiv#1{\href {http://arxiv.org/abs/#1} {{\tt arXiv:#1}}}
\def\mn@eprint@dblp#1{\href {http://dblp.uni-trier.de/rec/bibtex/#1.xml}
  {dblp:#1}}
\def\mn@eprint@#1:#2:#3:#4\@nil{\def\@tempa {#1}\def\@tempb {#2}\def\@tempc
  {#3}\ifx \@tempc \@empty \let \@tempc \@tempb \let \@tempb \@tempa \fi \ifx
  \@tempb \@empty \def\@tempb {arXiv}\fi \@ifundefined
  {mn@eprint@\@tempb}{\@tempb:\@tempc}{\expandafter \expandafter \csname
  mn@eprint@\@tempb\endcsname \expandafter{\@tempc}}}

\bibitem[\protect\citeauthoryear{{Aniano}, {Draine}, {Gordon}  \&
  {Sandstrom}}{{Aniano} et~al.}{2011}]{Aniano2011}
{Aniano} G.,  {Draine} B.~T.,  {Gordon} K.~D.,   {Sandstrom} K.,  2011, \mn@doi
  [\pasp] {10.1086/662219}, \href
  {https://ui.adsabs.harvard.edu/abs/2011PASP..123.1218A} {123, 1218}

\bibitem[\protect\citeauthoryear{{Aniyan} et~al.,}{{Aniyan}
  et~al.}{2018}]{Aniyan2018}
{Aniyan} S.,  et~al., 2018, \mn@doi [\mnras] {10.1093/mnras/sty310}, \href
  {https://ui.adsabs.harvard.edu/abs/2018MNRAS.476.1909A} {476, 1909}

\bibitem[\protect\citeauthoryear{{Bacchini}, {Fraternali}, {Iorio}  \&
  {Pezzulli}}{{Bacchini} et~al.}{2019a}]{Bacchini2019}
{Bacchini} C.,  {Fraternali} F.,  {Iorio} G.,   {Pezzulli} G.,  2019a, \mn@doi
  [\aap] {10.1051/0004-6361/201834382}, \href
  {https://ui.adsabs.harvard.edu/abs/2019A&A...622A..64B} {622, A64}

\bibitem[\protect\citeauthoryear{{Bacchini}, {Fraternali}, {Pezzulli},
  {Marasco}, {Iorio}  \& {Nipoti}}{{Bacchini} et~al.}{2019b}]{Bacchini2019a}
{Bacchini} C.,  {Fraternali} F.,  {Pezzulli} G.,  {Marasco} A.,  {Iorio} G.,
  {Nipoti} C.,  2019b, \mn@doi [\aap] {10.1051/0004-6361/201936559}, \href
  {https://ui.adsabs.harvard.edu/abs/2019A&A...632A.127B} {632, A127}

\bibitem[\protect\citeauthoryear{{Bacchini}, {Fraternali}, {Pezzulli}  \&
  {Marasco}}{{Bacchini} et~al.}{2020}]{Bacchini2020}
{Bacchini} C.,  {Fraternali} F.,  {Pezzulli} G.,   {Marasco} A.,  2020, \mn@doi
  [\aap] {10.1051/0004-6361/202038962}, \href
  {https://ui.adsabs.harvard.edu/abs/2020A&A...644A.125B} {644, A125}

\bibitem[\protect\citeauthoryear{{Banerjee}, {Jog}, {Brinks}  \&
  {Bagetakos}}{{Banerjee} et~al.}{2011}]{Banerjee2011}
{Banerjee} A.,  {Jog} C.~J.,  {Brinks} E.,   {Bagetakos} I.,  2011, \mn@doi
  [\mnras] {10.1111/j.1365-2966.2011.18745.x}, \href
  {https://ui.adsabs.harvard.edu/abs/2011MNRAS.415..687B} {415, 687}

\bibitem[\protect\citeauthoryear{{Begeman}, {Broeils}  \& {Sanders}}{{Begeman}
  et~al.}{1991}]{Begeman1991}
{Begeman} K.~G.,  {Broeils} A.~H.,   {Sanders} R.~H.,  1991, \mn@doi [\mnras]
  {10.1093/mnras/249.3.523}, \href
  {https://ui.adsabs.harvard.edu/abs/1991MNRAS.249..523B} {249, 523}

\bibitem[\protect\citeauthoryear{{Bergin} \& {Tafalla}}{{Bergin} \&
  {Tafalla}}{2007}]{Bergin2007}
{Bergin} E.~A.,  {Tafalla} M.,  2007, \mn@doi [\araa]
  {10.1146/annurev.astro.45.071206.100404}, \href
  {https://ui.adsabs.harvard.edu/abs/2007ARA&A..45..339B} {45, 339}

\bibitem[\protect\citeauthoryear{{Berkhuijsen}}{{Berkhuijsen}}{1977}]{Berkhuijsen1977}
{Berkhuijsen} E.~M.,  1977, \aap, \href
  {https://ui.adsabs.harvard.edu/abs/1977A&A....57....9B} {57, 9}

\bibitem[\protect\citeauthoryear{{Bigiel}, {Leroy}, {Walter}, {Brinks}, {de
  Blok}, {Madore}  \& {Thornley}}{{Bigiel} et~al.}{2008}]{Bigiel2008}
{Bigiel} F.,  {Leroy} A.,  {Walter} F.,  {Brinks} E.,  {de Blok} W.~J.~G.,
  {Madore} B.,   {Thornley} M.~D.,  2008, \mn@doi [\aj]
  {10.1088/0004-6256/136/6/2846}, \href
  {https://ui.adsabs.harvard.edu/abs/2008AJ....136.2846B} {136, 2846}

\bibitem[\protect\citeauthoryear{{Bolatto}, {Wolfire}  \& {Leroy}}{{Bolatto}
  et~al.}{2013}]{Bolatto2013}
{Bolatto} A.~D.,  {Wolfire} M.,   {Leroy} A.~K.,  2013, \mn@doi [\araa]
  {10.1146/annurev-astro-082812-140944}, \href
  {https://ui.adsabs.harvard.edu/abs/2013ARA&A..51..207B} {51, 207}

\bibitem[\protect\citeauthoryear{{Cald{\'u}-Primo}, {Schruba}, {Walter},
  {Leroy}, {Sandstrom}, {de Blok}, {Ianjamasimanana}  \&
  {Mogotsi}}{{Cald{\'u}-Primo} et~al.}{2013}]{CalduPrimo2013}
{Cald{\'u}-Primo} A.,  {Schruba} A.,  {Walter} F.,  {Leroy} A.,  {Sandstrom}
  K.,  {de Blok} W.~J.~G.,  {Ianjamasimanana} R.,   {Mogotsi} K.~M.,  2013,
  \mn@doi [\aj] {10.1088/0004-6256/146/6/150}, \href
  {https://ui.adsabs.harvard.edu/abs/2013AJ....146..150C} {146, 150}

\bibitem[\protect\citeauthoryear{{Cardelli}, {Clayton}  \& {Mathis}}{{Cardelli}
  et~al.}{1989}]{Cardelli1989}
{Cardelli} J.~A.,  {Clayton} G.~C.,   {Mathis} J.~S.,  1989, \mn@doi [\apj]
  {10.1086/167900}, \href
  {https://ui.adsabs.harvard.edu/abs/1989ApJ...345..245C} {345, 245}

\bibitem[\protect\citeauthoryear{{Dame}, {Hartmann}  \& {Thaddeus}}{{Dame}
  et~al.}{2001}]{Dame2001}
{Dame} T.~M.,  {Hartmann} D.,   {Thaddeus} P.,  2001, \mn@doi [\apj]
  {10.1086/318388}, \href
  {https://ui.adsabs.harvard.edu/abs/2001ApJ...547..792D} {547, 792}

\bibitem[\protect\citeauthoryear{{Di Teodoro} \& {Fraternali}}{{Di Teodoro} \&
  {Fraternali}}{2015}]{DiTeodoro2015}
{Di Teodoro} E.~M.,  {Fraternali} F.,  2015, \mn@doi [\mnras]
  {10.1093/mnras/stv1213}, \href
  {https://ui.adsabs.harvard.edu/abs/2015MNRAS.451.3021D} {451, 3021}

\bibitem[\protect\citeauthoryear{{Elmegreen}}{{Elmegreen}}{1997}]{Elmegreen1997}
{Elmegreen} B.~G.,  1997, in {Franco} J.,  {Terlevich} R.,   {Serrano} A.,
  eds,  Revista Mexicana de Astronomia y Astrofisica Conference Series Vol. 6,
  Revista Mexicana de Astronomia y Astrofisica Conference Series. p.~165

\bibitem[\protect\citeauthoryear{{Elmegreen} \& {Hunter}}{{Elmegreen} \&
  {Hunter}}{2015}]{Elmegreen2015}
{Elmegreen} B.~G.,  {Hunter} D.~A.,  2015, \mn@doi [\apj]
  {10.1088/0004-637X/805/2/145}, \href
  {https://ui.adsabs.harvard.edu/abs/2015ApJ...805..145E} {805, 145}

\bibitem[\protect\citeauthoryear{{Fierlinger}, {Burkert}, {Ntormousi},
  {Fierlinger}, {Schartmann}, {Ballone}, {Krause}  \& {Diehl}}{{Fierlinger}
  et~al.}{2016}]{Fierlinger2016}
{Fierlinger} K.~M.,  {Burkert} A.,  {Ntormousi} E.,  {Fierlinger} P.,
  {Schartmann} M.,  {Ballone} A.,  {Krause} M. G.~H.,   {Diehl} R.,  2016,
  \mn@doi [\mnras] {10.1093/mnras/stv2699}, \href
  {https://ui.adsabs.harvard.edu/abs/2016MNRAS.456..710F} {456, 710}

\bibitem[\protect\citeauthoryear{{Freedman}}{{Freedman}}{1984}]{Freedman1984}
{Freedman} W.~L.,  1984, PhD thesis, UNIVERSITY OF TORONTO (CANADA).

\bibitem[\protect\citeauthoryear{{Gao} \& {Solomon}}{{Gao} \&
  {Solomon}}{2004a}]{Gao2004}
{Gao} Y.,  {Solomon} P.~M.,  2004a, \mn@doi [\apjs] {10.1086/383003}, \href
  {https://ui.adsabs.harvard.edu/abs/2004ApJS..152...63G} {152, 63}

\bibitem[\protect\citeauthoryear{{Gao} \& {Solomon}}{{Gao} \&
  {Solomon}}{2004b}]{Gao2004a}
{Gao} Y.,  {Solomon} P.~M.,  2004b, \mn@doi [\apj] {10.1086/382999}, \href
  {https://ui.adsabs.harvard.edu/abs/2004ApJ...606..271G} {606, 271}

\bibitem[\protect\citeauthoryear{{Gil de Paz} et~al.,}{{Gil de Paz}
  et~al.}{2007}]{GildePaz2007}
{Gil de Paz} A.,  et~al., 2007, \mn@doi [\apjs] {10.1086/516636}, \href
  {https://ui.adsabs.harvard.edu/abs/2007ApJS..173..185G} {173, 185}

\bibitem[\protect\citeauthoryear{{Hao}, {Kennicutt}, {Johnson}, {Calzetti},
  {Dale}  \& {Moustakas}}{{Hao} et~al.}{2011}]{Hao2011}
{Hao} C.-N.,  {Kennicutt} R.~C.,  {Johnson} B.~D.,  {Calzetti} D.,  {Dale}
  D.~A.,   {Moustakas} J.,  2011, \mn@doi [\apj] {10.1088/0004-637X/741/2/124},
  \href {https://ui.adsabs.harvard.edu/abs/2011ApJ...741..124H} {741, 124}

\bibitem[\protect\citeauthoryear{{Hunter} et~al.,}{{Hunter}
  et~al.}{2012}]{Hunter2012}
{Hunter} D.~A.,  et~al., 2012, \mn@doi [\aj] {10.1088/0004-6256/144/5/134},
  \href {https://ui.adsabs.harvard.edu/abs/2012AJ....144..134H} {144, 134}

\bibitem[\protect\citeauthoryear{{Hunter}, {Elmegreen}  \& {Berger}}{{Hunter}
  et~al.}{2019}]{Hunter2019}
{Hunter} D.~A.,  {Elmegreen} B.~G.,   {Berger} C.~L.,  2019, \mn@doi [\aj]
  {10.3847/1538-3881/ab1e54}, \href
  {https://ui.adsabs.harvard.edu/abs/2019AJ....157..241H} {157, 241}

\bibitem[\protect\citeauthoryear{{Iorio}, {Fraternali}, {Nipoti}, {Di Teodoro},
  {Read}  \& {Battaglia}}{{Iorio} et~al.}{2017}]{Iorio2017}
{Iorio} G.,  {Fraternali} F.,  {Nipoti} C.,  {Di Teodoro} E.,  {Read} J.~I.,
  {Battaglia} G.,  2017, \mn@doi [\mnras] {10.1093/mnras/stw3285}, \href
  {https://ui.adsabs.harvard.edu/abs/2017MNRAS.466.4159I} {466, 4159}

\bibitem[\protect\citeauthoryear{{Kennicutt}}{{Kennicutt}}{1989}]{Kennicutt1989}
{Kennicutt} Robert~C. J.,  1989, \mn@doi [\apj] {10.1086/167834}, \href
  {https://ui.adsabs.harvard.edu/abs/1989ApJ...344..685K} {344, 685}

\bibitem[\protect\citeauthoryear{{Kennicutt}}{{Kennicutt}}{1998}]{Kennicutt1998}
{Kennicutt} Robert~C. J.,  1998, \mn@doi [\apj] {10.1086/305588}, \href
  {https://ui.adsabs.harvard.edu/abs/1998ApJ...498..541K} {498, 541}

\bibitem[\protect\citeauthoryear{{Kennicutt} Robert~C. et~al.,}{{Kennicutt}
  et~al.}{2003}]{Kennicutt2003}
{Kennicutt} Robert~C. J.,  et~al., 2003, \mn@doi [\pasp] {10.1086/376941},
  \href {https://ui.adsabs.harvard.edu/abs/2003PASP..115..928K} {115, 928}

\bibitem[\protect\citeauthoryear{{Kim}, {Kim}  \& {Ostriker}}{{Kim}
  et~al.}{2011}]{Kim2011}
{Kim} C.-G.,  {Kim} W.-T.,   {Ostriker} E.~C.,  2011, \mn@doi [\apj]
  {10.1088/0004-637X/743/1/25}, \href
  {https://ui.adsabs.harvard.edu/abs/2011ApJ...743...25K} {743, 25}

\bibitem[\protect\citeauthoryear{{Kregel}, {van der Kruit}  \& {de
  Grijs}}{{Kregel} et~al.}{2002}]{Kregel2002}
{Kregel} M.,  {van der Kruit} P.~C.,   {de Grijs} R.,  2002, \mn@doi [\mnras]
  {10.1046/j.1365-8711.2002.05556.x}, \href
  {https://ui.adsabs.harvard.edu/abs/2002MNRAS.334..646K} {334, 646}

\bibitem[\protect\citeauthoryear{{Krumholz}, {Dekel}  \& {McKee}}{{Krumholz}
  et~al.}{2012}]{Krumholz2012}
{Krumholz} M.~R.,  {Dekel} A.,   {McKee} C.~F.,  2012, \mn@doi [\apj]
  {10.1088/0004-637X/745/1/69}, \href
  {https://ui.adsabs.harvard.edu/abs/2012ApJ...745...69K} {745, 69}

\bibitem[\protect\citeauthoryear{{Leroy}, {Bolatto}, {Simon}  \&
  {Blitz}}{{Leroy} et~al.}{2005}]{Leroy2005}
{Leroy} A.,  {Bolatto} A.~D.,  {Simon} J.~D.,   {Blitz} L.,  2005, \mn@doi
  [\apj] {10.1086/429578}, \href
  {https://ui.adsabs.harvard.edu/abs/2005ApJ...625..763L} {625, 763}

\bibitem[\protect\citeauthoryear{{Leroy}, {Walter}, {Brinks}, {Bigiel}, {de
  Blok}, {Madore}  \& {Thornley}}{{Leroy} et~al.}{2008}]{Leroy2008}
{Leroy} A.~K.,  {Walter} F.,  {Brinks} E.,  {Bigiel} F.,  {de Blok} W.~J.~G.,
  {Madore} B.,   {Thornley} M.~D.,  2008, \mn@doi [\aj]
  {10.1088/0004-6256/136/6/2782}, \href
  {https://ui.adsabs.harvard.edu/abs/2008AJ....136.2782L} {136, 2782}

\bibitem[\protect\citeauthoryear{{Leroy} et~al.,}{{Leroy}
  et~al.}{2013}]{Leroy2013}
{Leroy} A.~K.,  et~al., 2013, \mn@doi [\aj] {10.1088/0004-6256/146/2/19}, \href
  {https://ui.adsabs.harvard.edu/abs/2013AJ....146...19L} {146, 19}

\bibitem[\protect\citeauthoryear{{Marasco}, {Fraternali}, {van der Hulst}  \&
  {Oosterloo}}{{Marasco} et~al.}{2017}]{Marasco2017}
{Marasco} A.,  {Fraternali} F.,  {van der Hulst} J.~M.,   {Oosterloo} T.,
  2017, \mn@doi [\aap] {10.1051/0004-6361/201731054}, \href
  {https://ui.adsabs.harvard.edu/abs/2017A&A...607A.106M} {607, A106}

\bibitem[\protect\citeauthoryear{{Mogotsi}, {de Blok}, {Cald{\'u}-Primo},
  {Walter}, {Ianjamasimanana}  \& {Leroy}}{{Mogotsi}
  et~al.}{2016}]{Mogotsi2016}
{Mogotsi} K.~M.,  {de Blok} W.~J.~G.,  {Cald{\'u}-Primo} A.,  {Walter} F.,
  {Ianjamasimanana} R.,   {Leroy} A.~K.,  2016, \mn@doi [\aj]
  {10.3847/0004-6256/151/1/15}, \href
  {https://ui.adsabs.harvard.edu/abs/2016AJ....151...15M} {151, 15}

\bibitem[\protect\citeauthoryear{{Mosenkov}, {Sotnikova}  \&
  {Reshetnikov}}{{Mosenkov} et~al.}{2010}]{Mosenkov2010a}
{Mosenkov} A.~V.,  {Sotnikova} N.~Y.,   {Reshetnikov} V.~P.,  2010, \mn@doi
  [\mnras] {10.1111/j.1365-2966.2009.15671.x}, \href
  {https://ui.adsabs.harvard.edu/abs/2010MNRAS.401..559M} {401, 559}

\bibitem[\protect\citeauthoryear{{Navarro}, {Frenk}  \& {White}}{{Navarro}
  et~al.}{1997}]{Navarro1997}
{Navarro} J.~F.,  {Frenk} C.~S.,   {White} S. D.~M.,  1997, \mn@doi [\apj]
  {10.1086/304888}, \href
  {https://ui.adsabs.harvard.edu/abs/1997ApJ...490..493N} {490, 493}

\bibitem[\protect\citeauthoryear{{Noordermeer} \& {van der
  Hulst}}{{Noordermeer} \& {van der Hulst}}{2007}]{Noordermeer2007}
{Noordermeer} E.,  {van der Hulst} J.~M.,  2007, \mn@doi [\mnras]
  {10.1111/j.1365-2966.2007.11532.x}, \href
  {https://ui.adsabs.harvard.edu/abs/2007MNRAS.376.1480N} {376, 1480}

\bibitem[\protect\citeauthoryear{{Oh}, {de Blok}, {Brinks}, {Walter}  \&
  {Kennicutt}}{{Oh} et~al.}{2011}]{Oh2011b}
{Oh} S.-H.,  {de Blok} W.~J.~G.,  {Brinks} E.,  {Walter} F.,   {Kennicutt}
  Robert~C. J.,  2011, \mn@doi [\aj] {10.1088/0004-6256/141/6/193}, \href
  {https://ui.adsabs.harvard.edu/abs/2011AJ....141..193O} {141, 193}

\bibitem[\protect\citeauthoryear{{Oh} et~al.,}{{Oh} et~al.}{2015}]{Oh2015}
{Oh} S.-H.,  et~al., 2015, \mn@doi [\aj] {10.1088/0004-6256/149/6/180}, \href
  {https://ui.adsabs.harvard.edu/abs/2015AJ....149..180O} {149, 180}

\bibitem[\protect\citeauthoryear{{Onodera} et~al.,}{{Onodera}
  et~al.}{2010}]{Onodera2010}
{Onodera} S.,  et~al., 2010, \mn@doi [\apjl] {10.1088/2041-8205/722/2/L127},
  \href {https://ui.adsabs.harvard.edu/abs/2010ApJ...722L.127O} {722, L127}

\bibitem[\protect\citeauthoryear{{Ostriker}, {McKee}  \& {Leroy}}{{Ostriker}
  et~al.}{2010}]{Ostriker2010}
{Ostriker} E.~C.,  {McKee} C.~F.,   {Leroy} A.~K.,  2010, \mn@doi [\apj]
  {10.1088/0004-637X/721/2/975}, \href
  {https://ui.adsabs.harvard.edu/abs/2010ApJ...721..975O} {721, 975}

\bibitem[\protect\citeauthoryear{{Pastrav}, {Popescu}, {Tuffs}  \&
  {Sansom}}{{Pastrav} et~al.}{2013}]{Pastrav2013}
{Pastrav} B.~A.,  {Popescu} C.~C.,  {Tuffs} R.~J.,   {Sansom} A.~E.,  2013,
  \mn@doi [\aap] {10.1051/0004-6361/201220962}, \href
  {https://ui.adsabs.harvard.edu/abs/2013A&A...553A..80P} {553, A80}

\bibitem[\protect\citeauthoryear{{Rohlfs}}{{Rohlfs}}{1977}]{Rohlfs1977}
{Rohlfs} K.,  1977, {Lectures on density wave theory}.
 Vol. 69, \mn@doi{10.1007/3-540-08448-7, }

\bibitem[\protect\citeauthoryear{{Roychowdhury}, {Chengalur}  \&
  {Shi}}{{Roychowdhury} et~al.}{2017}]{Roychowdhury2017}
{Roychowdhury} S.,  {Chengalur} J.~N.,   {Shi} Y.,  2017, \mn@doi [\aap]
  {10.1051/0004-6361/201731083}, \href
  {https://ui.adsabs.harvard.edu/abs/2017A&A...608A..24R} {608, A24}

\bibitem[\protect\citeauthoryear{{Sandstrom} et~al.,}{{Sandstrom}
  et~al.}{2013}]{Sandstrom2013}
{Sandstrom} K.~M.,  et~al., 2013, \mn@doi [\apj] {10.1088/0004-637X/777/1/5},
  \href {https://ui.adsabs.harvard.edu/abs/2013ApJ...777....5S} {777, 5}

\bibitem[\protect\citeauthoryear{{Schaye} \& {Dalla Vecchia}}{{Schaye} \&
  {Dalla Vecchia}}{2008}]{Schaye2008}
{Schaye} J.,  {Dalla Vecchia} C.,  2008, \mn@doi [\mnras]
  {10.1111/j.1365-2966.2007.12639.x}, \href
  {https://ui.adsabs.harvard.edu/abs/2008MNRAS.383.1210S} {383, 1210}

\bibitem[\protect\citeauthoryear{{Schlafly} \& {Finkbeiner}}{{Schlafly} \&
  {Finkbeiner}}{2011}]{Schlafly2011}
{Schlafly} E.~F.,  {Finkbeiner} D.~P.,  2011, \mn@doi [\apj]
  {10.1088/0004-637X/737/2/103}, \href
  {https://ui.adsabs.harvard.edu/abs/2011ApJ...737..103S} {737, 103}

\bibitem[\protect\citeauthoryear{{Schmidt}}{{Schmidt}}{1959}]{Schmidt1959}
{Schmidt} M.,  1959, \mn@doi [\apj] {10.1086/146614}, \href
  {https://ui.adsabs.harvard.edu/abs/1959ApJ...129..243S} {129, 243}

\bibitem[\protect\citeauthoryear{{Semenov}, {Kravtsov}  \& {Gnedin}}{{Semenov}
  et~al.}{2018}]{Semenov2018}
{Semenov} V.~A.,  {Kravtsov} A.~V.,   {Gnedin} N.~Y.,  2018, \mn@doi [\apj]
  {10.3847/1538-4357/aac6eb}, \href
  {https://ui.adsabs.harvard.edu/abs/2018ApJ...861....4S} {861, 4}

\bibitem[\protect\citeauthoryear{{Shi}, {Helou}, {Yan}, {Armus}, {Wu},
  {Papovich}  \& {Stierwalt}}{{Shi} et~al.}{2011}]{Shi2011}
{Shi} Y.,  {Helou} G.,  {Yan} L.,  {Armus} L.,  {Wu} Y.,  {Papovich} C.,
  {Stierwalt} S.,  2011, \mn@doi [\apj] {10.1088/0004-637X/733/2/87}, \href
  {https://ui.adsabs.harvard.edu/abs/2011ApJ...733...87S} {733, 87}

\bibitem[\protect\citeauthoryear{{Shi}, {Wang}, {Zhang}, {Gao}, {Hao}, {Xia}
  \& {Gu}}{{Shi} et~al.}{2016}]{Shi2016}
{Shi} Y.,  {Wang} J.,  {Zhang} Z.-Y.,  {Gao} Y.,  {Hao} C.-N.,  {Xia} X.-Y.,
  {Gu} Q.,  2016, \mn@doi [Nature Communications] {10.1038/ncomms13789}, \href
  {https://ui.adsabs.harvard.edu/abs/2016NatCo...713789S} {7, 13789}

\bibitem[\protect\citeauthoryear{{Shi} et~al.,}{{Shi} et~al.}{2018}]{Shi2018}
{Shi} Y.,  et~al., 2018, \mn@doi [\apj] {10.3847/1538-4357/aaa3e6}, \href
  {https://ui.adsabs.harvard.edu/abs/2018ApJ...853..149S} {853, 149}

\bibitem[\protect\citeauthoryear{{Shi}, {Zhang}, {Wang}, {Chen}, {Gu}, {Yu}  \&
  {Li}}{{Shi} et~al.}{2021}]{Shi2021}
{Shi} Y.,  {Zhang} Z.-Y.,  {Wang} J.,  {Chen} J.,  {Gu} Q.,  {Yu} X.,   {Li}
  S.,  2021, \mn@doi [\apj] {10.3847/1538-4357/abd777}, \href
  {https://ui.adsabs.harvard.edu/abs/2021ApJ...909...20S} {909, 20}

\bibitem[\protect\citeauthoryear{{Silk}}{{Silk}}{1997}]{Silk1997}
{Silk} J.,  1997, \mn@doi [\apj] {10.1086/304073}, \href
  {https://ui.adsabs.harvard.edu/abs/1997ApJ...481..703S} {481, 703}

\bibitem[\protect\citeauthoryear{{Walter}, {Brinks}, {de Blok}, {Bigiel},
  {Kennicutt}, {Thornley}  \& {Leroy}}{{Walter} et~al.}{2008}]{Walter2008}
{Walter} F.,  {Brinks} E.,  {de Blok} W.~J.~G.,  {Bigiel} F.,  {Kennicutt}
  Robert~C. J.,  {Thornley} M.~D.,   {Leroy} A.,  2008, \mn@doi [\aj]
  {10.1088/0004-6256/136/6/2563}, \href
  {https://ui.adsabs.harvard.edu/abs/2008AJ....136.2563W} {136, 2563}

\bibitem[\protect\citeauthoryear{{Wang}, {Yang}, {Zhang}, {Fang}, {Shi}, {Liu},
  {Li}  \& {Li}}{{Wang} et~al.}{2020}]{Wang2020}
{Wang} J.,  {Yang} K.,  {Zhang} Z.-Y.,  {Fang} M.,  {Shi} Y.,  {Liu} S.,  {Li}
  J.,   {Li} F.,  2020, \mn@doi [\mnras] {10.1093/mnrasl/slaa150}, \href
  {https://ui.adsabs.harvard.edu/abs/2020MNRAS.499L..26W} {499, L26}

\bibitem[\protect\citeauthoryear{{Wyder} et~al.,}{{Wyder}
  et~al.}{2009}]{Wyder2009}
{Wyder} T.~K.,  et~al., 2009, \mn@doi [\apj] {10.1088/0004-637X/696/2/1834},
  \href {https://ui.adsabs.harvard.edu/abs/2009ApJ...696.1834W} {696, 1834}

\bibitem[\protect\citeauthoryear{{Yim}, {Wong}, {Howk}  \& {van der
  Hulst}}{{Yim} et~al.}{2011}]{Yim2011}
{Yim} K.,  {Wong} T.,  {Howk} J.~C.,   {van der Hulst} J.~M.,  2011, \mn@doi
  [\aj] {10.1088/0004-6256/141/2/48}, \href
  {https://ui.adsabs.harvard.edu/abs/2011AJ....141...48Y} {141, 48}

\bibitem[\protect\citeauthoryear{{Yim}, {Wong}, {Xue}, {Rand}, {Rosolowsky},
  {van der Hulst}, {Benjamin}  \& {Murphy}}{{Yim} et~al.}{2014}]{Yim2014}
{Yim} K.,  {Wong} T.,  {Xue} R.,  {Rand} R.~J.,  {Rosolowsky} E.,  {van der
  Hulst} J.~M.,  {Benjamin} R.,   {Murphy} E.~J.,  2014, \mn@doi [\aj]
  {10.1088/0004-6256/148/6/127}, \href
  {https://ui.adsabs.harvard.edu/abs/2014AJ....148..127Y} {148, 127}

\bibitem[\protect\citeauthoryear{{Yim}, {Wong}, {Rand}  \& {Schinnerer}}{{Yim}
  et~al.}{2020}]{Yim2020}
{Yim} K.,  {Wong} T.,  {Rand} R.~J.,   {Schinnerer} E.,  2020, \mn@doi [\mnras]
  {10.1093/mnras/staa1020}, \href
  {https://ui.adsabs.harvard.edu/abs/2020MNRAS.494.4558Y} {494, 4558}

\bibitem[\protect\citeauthoryear{{Zhou}, {Wu}, {Zhou}  \& {Ma}}{{Zhou}
  et~al.}{2018}]{Zhou2018}
{Zhou} Z.,  {Wu} H.,  {Zhou} X.,   {Ma} J.,  2018, \mn@doi [\pasp]
  {10.1088/1538-3873/aad407}, \href
  {https://ui.adsabs.harvard.edu/abs/2018PASP..130i4101Z} {130, 094101}

\bibitem[\protect\citeauthoryear{{de Blok}, {Walter}, {Brinks}, {Trachternach},
  {Oh}  \& {Kennicutt}}{{de Blok} et~al.}{2008}]{deBlok2008}
{de Blok} W.~J.~G.,  {Walter} F.,  {Brinks} E.,  {Trachternach} C.,  {Oh}
  S.~H.,   {Kennicutt} R.~C. J.,  2008, \mn@doi [\aj]
  {10.1088/0004-6256/136/6/2648}, \href
  {https://ui.adsabs.harvard.edu/abs/2008AJ....136.2648D} {136, 2648}

\bibitem[\protect\citeauthoryear{{de Grijs} \& {Peletier}}{{de Grijs} \&
  {Peletier}}{1997}]{deGrijs1997}
{de Grijs} R.,  {Peletier} R.~F.,  1997, \aap, \href
  {https://ui.adsabs.harvard.edu/abs/1997A&A...320L..21D} {320, L21}

\bibitem[\protect\citeauthoryear{{van der Kruit} \& {Searle}}{{van der Kruit}
  \& {Searle}}{1981}]{vanderKruit1981a}
{van der Kruit} P.~C.,  {Searle} L.,  1981, \aap, \href
  {https://ui.adsabs.harvard.edu/abs/1981A&A....95..105V} {95, 105}

\makeatother
\end{thebibliography}

\appendix

\section{Volumetric ES and KS law based on azimuthally-averaged quanities}

\begin{figure*}
  \centering
  \subfloat{\includegraphics[width=0.45\textwidth]{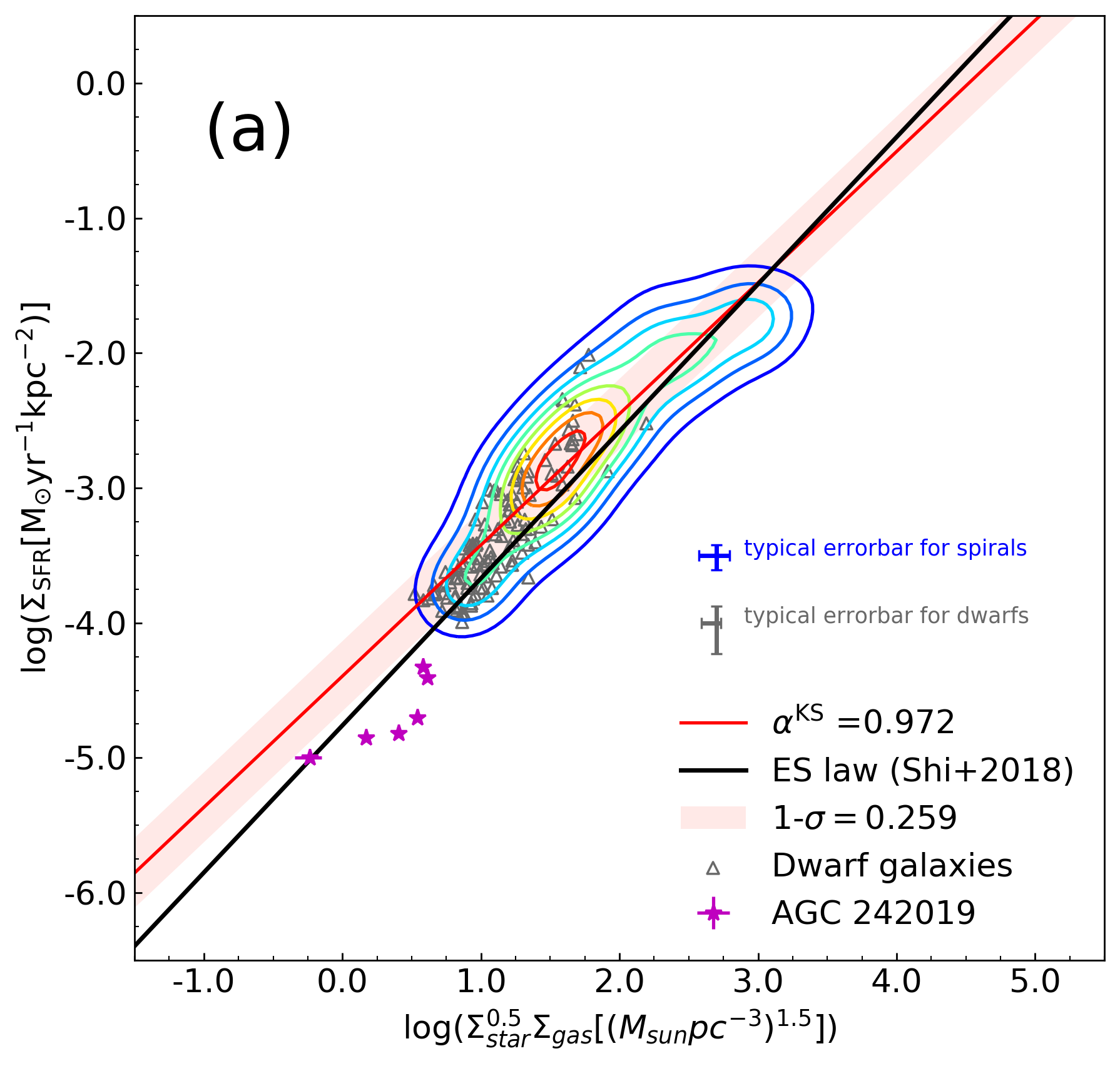}}\quad
  \subfloat{\includegraphics[width=0.45\textwidth]{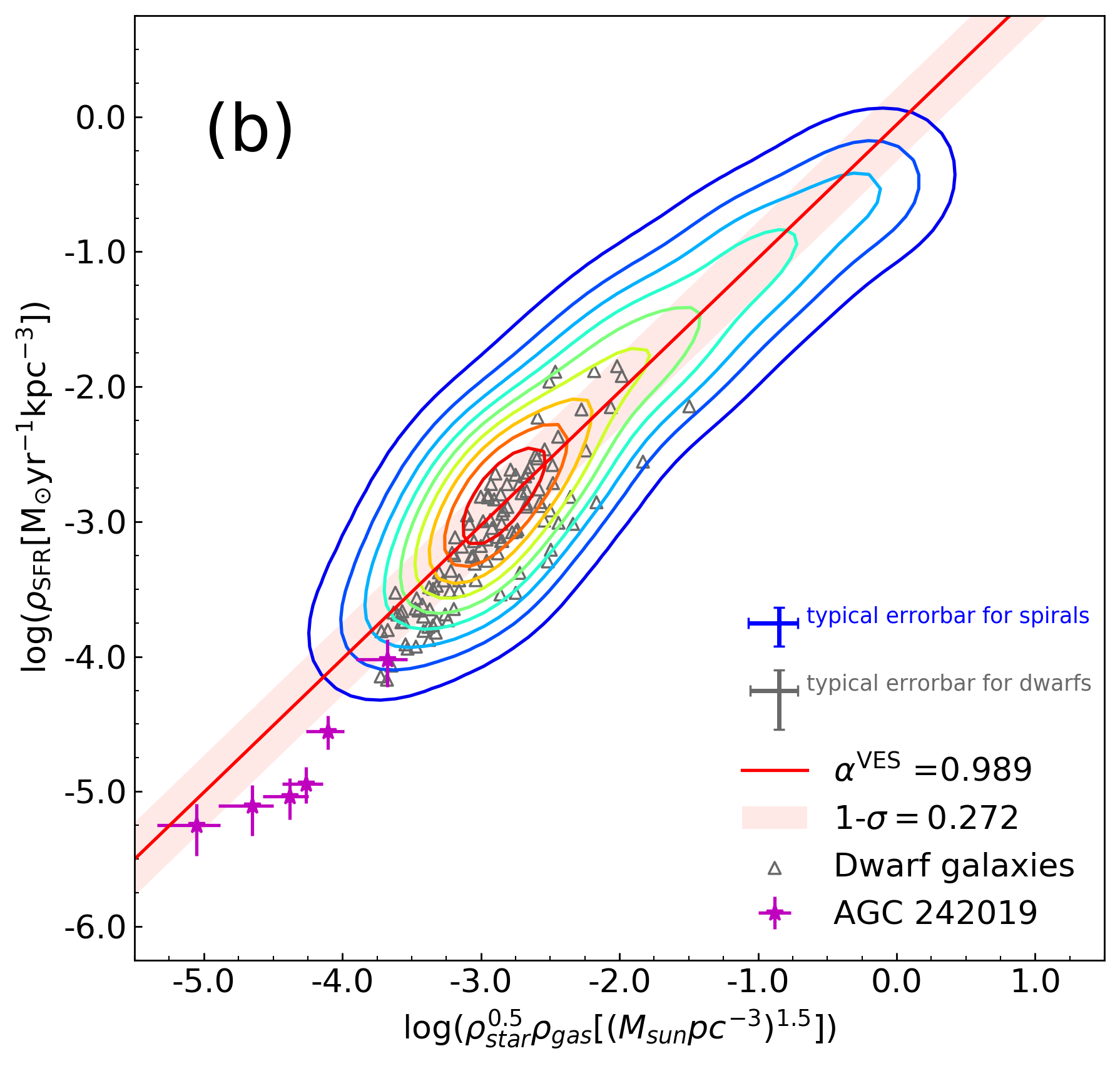}}\\
  \subfloat{\includegraphics[width=0.45\textwidth]{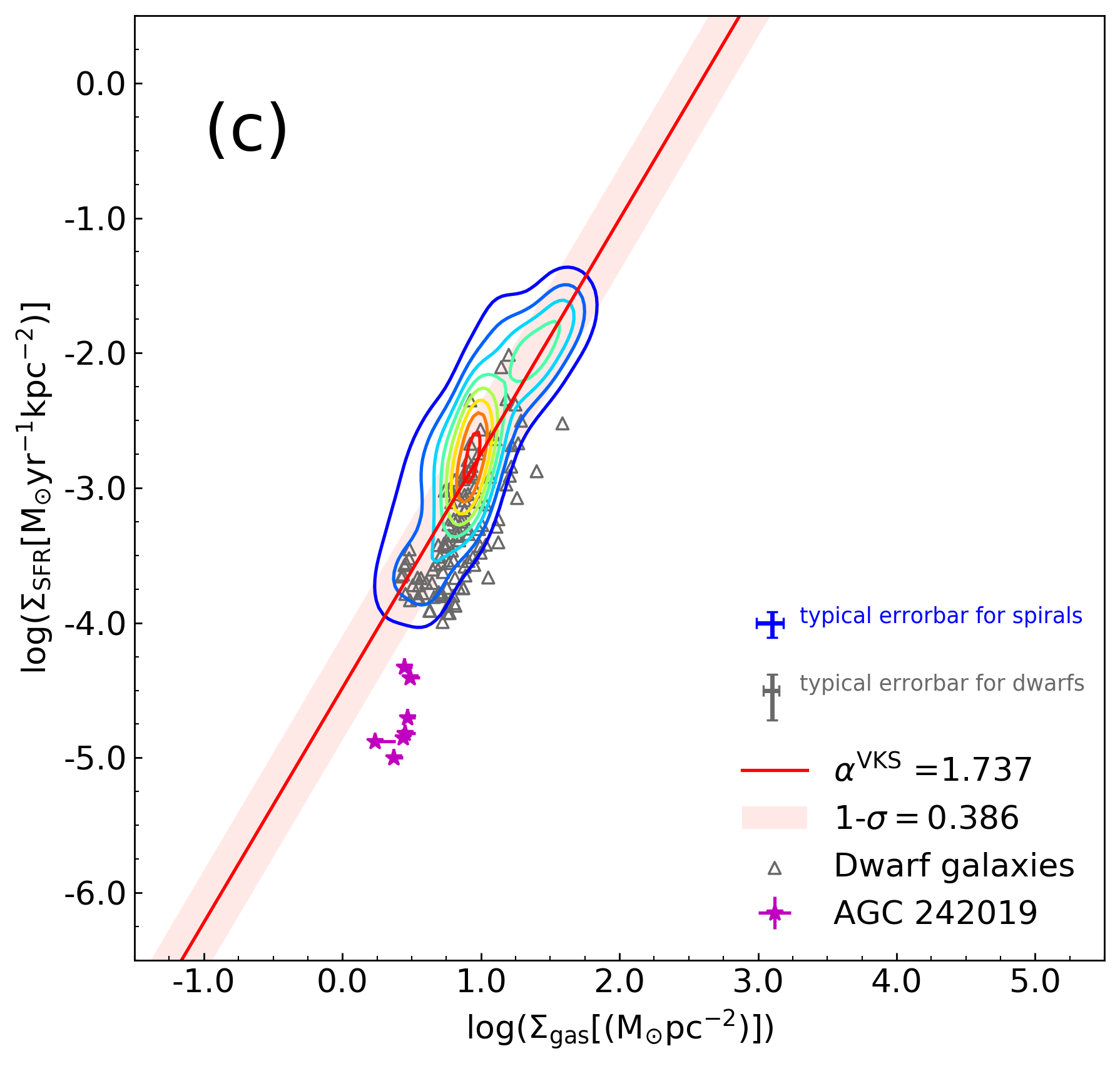}}\quad
  \subfloat{\includegraphics[width=0.45\textwidth]{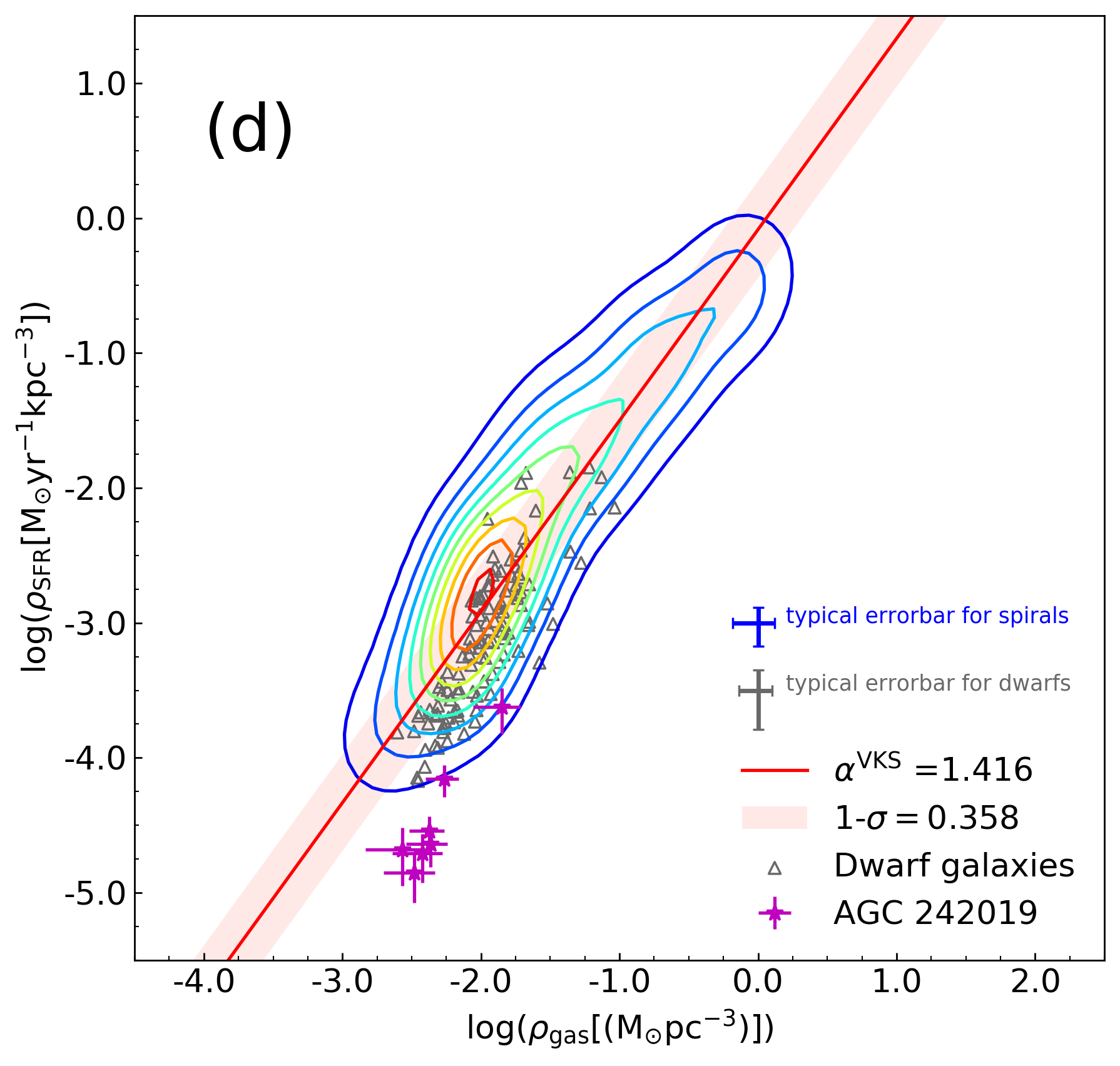}}
  \caption{Panel (a) and (b) show the ES law in surface densities and volume densities based on azimuthally-averaged quantities. Panel (c) and (d) show the KS law in surface densities and volume densities based on azimuthally-averaged quantities. Colored contours show the number density of the data points of spiral galaxies. Grey triangles represent dwarf galaxies and magenta stars with errorbars represent AGC~242019. In each panel, the typical errorbars of spiral galaxies and dwarf galaxies are plotted in blue and grey errorbars, respectively. The red line represents the best-fit model for our sample and the light-red region shows rms scatter bounds for the best-fit model. The black line in panel (a) is the best-fit model of surface-density ES law in \citet{Shi2018}.}
  \label{fig.A1}
\end{figure*}

Figure~\ref{fig.A1} shows the ES and KS law based on azimuthally-averaged quantities. Both surface-density and volumetric ES law have smaller rms scatter than surface-density and volumetric KS law, which is also found in our main result. The slope and rms scatter of volumetric ES law are 0.99 and 0.27,respectively. The slope is slightly larger than that of the volumetric ES law based on local measurements. The slope and rms scatter of the volumetric KS law are 1.42 and 0.36, respectively. The slope is also slightly larger than the volumetric KS law based on local measurements but much smaller than that ($\alpha^{\rm VKS}\sim1.91$) obtained in \citet{Bacchini2019}, in which the volumetric KS law is based on azimuthally-averaged quantities. The analysis based on azimuthally-averaged quantities does not change the main result of our work.


\bsp	
\label{lastpage}
\end{document}